\documentclass[notitlepage,aps,pra,twoside,superscriptaddress,showkeys,nofootinbib]{revtex4-1}
\usepackage{graphicx}
\usepackage{amsmath,amsfonts,amssymb,amsthm,amsbsy,mathtools}
\usepackage{amsmath} 
\DeclareMathOperator{\tr}{tr}

\usepackage{bbold}
\usepackage{epic}
\usepackage{eepic}
\usepackage{mathrsfs}
\usepackage[all]{xy}
\usepackage{lmodern}
\usepackage[T1]{fontenc}
\usepackage[utf8]{inputenc}
\usepackage{color}
\usepackage{xr}
\usepackage{blindtext}
\usepackage[utf8]{inputenc}
\usepackage[T1]{fontenc}
\usepackage{amssymb} \usepackage{color,graphicx} \usepackage{amsmath}
\usepackage{amsbsy} \usepackage{amsthm} 
\usepackage{bm} \usepackage{float} 
\usepackage{placeins,cals}
\usepackage{setspace,ulem}
\usepackage[colorlinks=true,allcolors=blue!80!black]{hyperref}
\usepackage{xfrac}
\usepackage{amsfonts}
\usepackage{latexsym}
\usepackage{amsmath,amssymb}
\usepackage{mathrsfs}
\usepackage{color,xcolor}
\usepackage{slashed}
\usepackage{url}
\usepackage{empheq}
\usepackage{textcomp}
\usepackage{dcolumn}
\usepackage{ulem}
\usepackage{subfigure}
\usepackage{amsfonts}
\usepackage{epsfig}
\usepackage{bbm}
\usepackage{tabularx}
\usepackage{multirow}
\usepackage{mathtools}
\usepackage{xfrac}
\usepackage{epsfig}
\usepackage{amsfonts}
\usepackage{latexsym}
\usepackage{mathrsfs}
\usepackage{hyperref}
\usepackage{setspace}
\usepackage{color}
\usepackage{bm}
\usepackage{orcidlink}
\usepackage{isotope}
\usepackage{braket}
\def \x{\mathbf{ x}}

\def \K{\mathbf{ K}}
\def \k{\mathbf{ k}}
\def \p{\mathbf{ p}}
\def \q{\mathbf{ q}}

\def \l{\left}
\def \r{\right}
\def\beq{\begin{equation}}
	\def\eeq{\end{equation}}
\def\bea{\begin{eqnarray}}
	\def\eea{\end{eqnarray}}
\usepackage{mhchem}

\def\k{\textbf{k}}

\renewcommand\Re{\operatorname{Re }}
\renewcommand\Im{\operatorname{Im }}

\allowdisplaybreaks
\makeatletter
\usepackage{cancel}
\usepackage{color}

\begin{document}
	
	\title{ Microscopic QED origin of spin entanglement}

	\author{M. Zarei\,\orcidlink{0000-0001-7744-2817}}
	\email{m.zarei@iut.ac.ir}
	\affiliation{Department of Physics, Isfahan University of Technology, 84156-83111 Isfahan, Iran}
	\affiliation{Quantum Technology Research Group, Isfahan University of Technology, Isfahan 84156-83111, Iran}

	\begin{abstract}
		
		We study effective spin interactions arising from quantum electrodynamics (QED) scattering between localized fermionic spins. By integrating out photon and mediator fields, the dynamics reduce to an effective spin Hamiltonian.
		For two qubits in the nonrelativistic regime, the resulting interaction takes a tensor dipolar form with an asymptotic decay proportional to \(R^{-3}\). We obtain analytical expressions for the entanglement negativity, highlighting its dependence on coupling strength and spatial configuration.
		We then examine a setup in which two bath spins interact via a sequential exchange with an intermediate fermionic mediator. At the perturbative order considered, the mediator remains unentangled and induces an effective bath--bath interaction with stronger spatial suppression than in the photon-mediated case.
			Extending the construction to an \(N\)-spin setting yields an effective interaction network mediated by virtual exchange processes, which can support the generation of multipartite entanglement across the system.

	\end{abstract}
	
	\maketitle

	\section{Introduction}
	
	 Quantum entanglement plays a central role in quantum information science, underpinning applications ranging from quantum computation and communication to precision metrology \cite{Horodecki:2009zz,Streltsov_2017,Wu_2021,Erhard_2020,Giovannetti_2011,manenti2023quantum,bertlmann2023modern}. Its controlled generation and preservation in increasingly complex systems is one of the main ingredients behind the expected advantages of quantum technologies over classical ones \cite{nielsen00}. In recent years, experimental progress has made it possible to prepare and manipulate entangled states across a wide variety of platforms, including trapped ions, superconducting circuits, neutral atoms, solid-state spin systems, and photonic architectures \cite{HAFFNER_2008,Anderlini_2007,O_Brien_2009,Bialczak_2010,Kjaergaard_2020,Flamini_2018,Henriet_2020}.

	 Depending on the physical setting, entanglement is typically generated through a few broad mechanisms. One common approach is gate-based control, where sequences of universal two-qubit operations such as CNOT or controlled-phase gates are used to build entangled states in a digitally programmable way. This framework underlies most quantum computing architectures and has been demonstrated in trapped ions \cite{Sackett:2000qkl,Leibfried:2003rkv,Blatt:2008jzf}, superconducting circuits \cite{McDermott2005,DiCarlo:2009alf,Barends:2014zey}, neutral atom systems using Rydberg interactions \cite{Urban:2008pgu,Levine:2019zfq}, and solid-state spin qubits in silicon devices \cite{Veldhorst:2014izh,Zajac:2017fee,Watson:2018lbm,He2019ATG}.
	 Another route is provided by Hamiltonian-driven dynamics, where entanglement emerges naturally under the intrinsic evolution of interacting systems. Here, initially separable states become entangled through native couplings such as exchange interactions, Ising-type terms, or dipole–dipole interactions. Examples include semiconductor quantum dots with tunable exchange coupling \cite{Petta:2005dew,Shulman:2012kke}, Rydberg atom arrays with effective Ising interactions \cite{Bernien:2017ubn}, and various solid-state spin platforms where dipolar or exchange couplings are used directly to generate entanglement \cite{Dolde:2012jur}.
	  A third class of protocols relies on measurement backaction or engineered dissipation. In these cases, entanglement is produced either probabilistically or as a steady-state property of the dynamics, often through repeated measurements or coupling to tailored reservoirs \cite{Cabrillo:1998jy,Browne:2003fuz,Beige:2000jin,Plenio:2002mog,Hensen:2015ccp}.
	 
	 At the same time, unavoidable coupling to the environment leads to decoherence, which gradually destroys quantum correlations and places limits on scalable quantum technologies \cite{Zurek:1991vd,Joos:1984uk,Braun_2001}. While open-system approaches such as master equations provide an effective framework to describe this process \cite{Breuer2002,Gisin1996}, they often obscure the connection between microscopic interactions and the resulting entanglement dynamics. In particular, the emergence of entanglement from underlying physical processes is usually encoded only indirectly through phenomenological parameters.
	 This separation becomes especially apparent in phenomena such as entanglement sudden death, where nonlocal correlations vanish in finite time even though local coherence decays smoothly \cite{Yu_2004,Dodd2004,Almeida:2007jib}. Despite significant progress, a fully microscopic understanding of how entanglement is generated and evolves in interacting quantum systems is still incomplete. In most practical treatments, effective spin Hamiltonians or noise models are introduced at a coarse-grained level, leaving the underlying physical origin of the couplings implicit.
	 
	 In this work, we construct effective spin Hamiltonians starting directly from QED scattering processes. We first analyze the two-qubit case, where photon exchange leads to a tensor spin–spin interaction with a dipolar structure. We then extend the framework to three qubits, where a single mediator couples two bath spins and generates both pairwise interactions and genuine three-body exchange terms. The approach generalizes naturally to an $N$-qubit network, producing a class of geometry-dependent many-body spin models. By varying the spatial arrangement and relative orientations of the scattering centers, the effective couplings can be tuned between Ising-, XX-, and Heisenberg-like forms.
	 
	 Using this microscopic Hamiltonian, we derive analytical expressions for the entanglement dynamics and characterize bipartite correlations via the negativity. We find that entanglement between bath spins is generated through mediator-induced interactions, while the mediator itself remains separable at the perturbative order considered. In this sense, it acts as a virtual channel that transfers correlations without becoming entangled with the system.
	 
	 The remainder of the paper is organized as follows. In Sec. II, we review a dynamical framework for entanglement generation in interacting qubit systems. In Sec. III and Appendix A, we study entanglement generated by photon exchange between two qubits. Section IV and Appendix B extend the analysis to two bath qubits interacting through a mediator qubit $A$. Finally, in Sec. V, we generalize the construction to an $N$-qubit system coupled via a common mediator. Unless stated otherwise, we use natural units with $\hbar = c = 1$.

	\section{Dynamical entanglement equation}

	Seminal works such as those by Konrad et al. \cite{Konrad2008}, who established a universal factorization law for concurrence under one sided channels, and Życzkowski et al. \cite{Zyczkowski2002}, who explored entanglement decay and revivals in discrete time dynamics, have significantly advanced our ability to understand entanglement dynamics. These approaches revealed that, under specific conditions, the evolution of entanglement can be reduced to simpler expressions dependent on the channel’s effect on a maximally entangled state. 
	
	However, existing models are often limited to specific two qubit systems and without a microscopic description of the underlying interaction mechanisms. 
	Here, we propose a dynamical equation derived from microscopic QED calculations to systematically study the creation of entanglement. Our approach models qubit interactions as fermionic systems coupled via photon propagator.

	The quantum Boltzman equation (QBE) formalism was originally formulated to presents a comprehensive technique for modeling the behavior of neutrinos that undergo flavor mixing while interacting with a medium \cite{Sigl:1993ctk} and also the time evolution of the intensity and polarization of cosmic microwave background (CMB) photons~\cite{Kosowsky:1994cy,Bavarsad:2009hm,Bartolo:2018igk,Bartolo:2019eac,Hoseinpour:2020hic} and quantum systems \cite{Sharifian:2023jem,Manshouri:2025nqc}. The mathematical explanation of the QBE is made simpler and more useful by the so called Born-Markov approximation. Under the Born approximation, the system interacts only weakly with its environment, which is assumed to be much larger than the system itself \cite{Zarei:2021dpb}. The QBE enables momentum resolved treatment and naturally incorporates propagator level structure.
	
	We have extended this formalism to investigate entanglement dynamics between two quantum systems, $A$ and $B$, providing a generalized technique for their correlated evolution \cite{Zarei:2025nnm}.
	To study the evolution of entanglement, this modified QBE is applied to the $4\times 4$ density matrix $\rho_{IJ}$
	\begin{align} 
		\left[(2\pi)^3\delta^3(0)\right]^2\dot{\rho}_{IJ}(\mathbf{k},t) =i\left\langle\big[\hat{H}_{\textrm{int}}(0),\hat{\mathcal{D}}_{IJ}(\mathbf{k})\big]\right\rangle -\int_{0}^{t}\hspace{-2mm}ds\left\langle\Big[\hat{H}_{\textrm{int}}(s),\big[\hat{H}^\dagger_{\textrm{int}}(0),\hat{\mathcal{D}}_{IJ}(\mathbf{k})\big]\Big]\right\rangle~,\label{QBE1}
	\end{align}
	where the capital indices are $I,J=1,\dots,4$ and $\hat{\mathcal{D}}_{IJ}$ is given by the tensor product of the number densities of systems $A$ and $B$, and $[(2\pi)^3\delta^3(0)]^2$ factors correspond to momentum space volume normalization.
	In this formalism, the expectation value of an operator $\hat{O}$ is given by
	\begin{equation} \label{exp1} \langle \hat{O}(\mathbf{k}) \rangle = \mathrm{tr}[\rho \hat{O}(\mathbf{k})] =\int d \p\left\langle\mathbf{p}\middle| \hat{\rho} \hat{O}(\mathbf{k}) \middle| \mathbf{p} \right\rangle~,
	\end{equation}
	with $d\mathbf{p} = d^3p/(2\pi)^3$ and the density operator is defined as
	\begin{equation}
		\hat{\rho}=\int d\p\rho_{IJ}(\mathbf{p})\hat{\mathcal{D}}_{IJ}(\mathbf{p})~,
	\end{equation}
	The interaction Hamiltonian 
	$H_{\textrm{int}}$ governs the scattering process of system 
	$A$ from system 
	$B$, and is defined via the S-matrix formalism, as will be illustrated in the examples presented in the subsequent sections.
	In Eq.~\eqref{QBE1}, the first term on the right-hand side represents forward scattering, where the momenta remain unchanged but phase shifts alter the quantum coherence. The second term, describes collision or decay processes, which induce decoherence and disentanglement. 
		In this work, we focus on investigating the dynamics of entanglement generation in a system consisting of two interacting qubits by photon exchange. In the subsequent sections, this formalism is also extended to the case in which qubit \(A\) interacts with \(N\) qubits \(B_1, \ldots, B_N\).

	\section{Entanglement generation between two qubits due to photon propagator }
	
	In the QED description used here, the two qubits are represented as localized
	wavepackets of a Dirac spinor field $\psi(x)$ coupled to the quantized
	electromagnetic field $A_\mu(x)$ field.
	The qubits follow prescribed worldlines $\bar x(\tau)$ and $\bar x'(\tau')$, and
	their finite spatial localization is incorporated through smearing functions
	centered on these trajectories. At the microscopic level, the effective interaction
	Hamiltonian describing the interaction two fermions via photon exchange is
	\begin{equation}
		\hat H_{\mathrm{eff}}
		= VV'q^2 \int d\tau\, d\tau'\, d^{3}x\, d^{4}x'\,
		A_{\mu}(x)\, A_{\nu}(x')\,
		\bar{\psi}^{-}(x)\gamma^{\mu}\psi^{+}(x)\,
		\delta_{\sigma_0}^{4}\!\left(x-\bar x(\tau)\right)\,
		\bar{\psi}^{-}(x')\gamma^{\nu}\psi^{+}(x')\,
		\delta_{\sigma_0'}^{4}\!\left(x'-\bar x'(\tau')\right)~,
		\label{H_int_model}
	\end{equation}
	where $A_\mu(x)$ is the electromagnetic four-potential, $\psi(x)$ is the Dirac
	spinor associated with the matter degrees of freedom, $q$ denotes the electric charge of qubits and the bilinear
	$\bar{\psi}\gamma^\mu\psi$ represents the local matter current. The functions
	$\delta_{\sigma_0}^{4}$ and $\delta_{\sigma_0'}^{4}$ encode the spacetime width
	of the qubit wavepackets~\cite{Breuer2002}.
	Wick contracting of the photon fields in the time-ordered evolution operator
	generated by $\hat H_{\mathrm{int}}$ yields the photon Feynman propagator as
	\begin{equation}
		P_F^{\mu\nu}(x-y) = \langle 0 | T \{ A^\mu(x) A^\nu(y) \} | 0 \rangle
		= -i\, \eta^{\mu\nu}\, P_F(x-y)~,
	\end{equation}
	with the Fourier transform as
	\begin{equation}
		P_F^{\mu\nu}(x) = \int \frac{d^4 K}{(2\pi)^4} \, e^{-i K \cdot x}\,
		P_F^{\mu\nu}(K)
		= \int \frac{d^4 K}{(2\pi)^4} \, e^{-i K \cdot x} \,
		\frac{-i \, \eta^{\mu\nu}}{K^2 + i\epsilon}~.
	\end{equation}
After inserting $	\hat H_{\mathrm{eff}}$ in forward scattering term of \eqref{QBE1} and performing straightforward calculations (see Appendix A for details),
we find the time evolution of $\rho_{IJ}$
of two localized spin-$1/2$ particles $A$ and $B$ separated by a fixed distance $\mathbf R$ as
 (\eqref{rhoReq2})
\begin{align}
	\dot{\rho}_{IJ}(\mathbf R)
	&=
	i\Gamma_{SS}(R)
	\Big[
	(\boldsymbol{\sigma}\rho)_{ij}\!\cdot\!(\boldsymbol{\sigma}\rho)_{kl}
	-
	3(\boldsymbol{\sigma}\rho)_{ij}\!\cdot\!\hat{\mathbf R}\,
	(\boldsymbol{\sigma}\rho)_{kl}\!\cdot\!\hat{\mathbf R}
	-
	(\rho\boldsymbol{\sigma})_{ij}\!\cdot\!(\rho\boldsymbol{\sigma})_{kl}
	+
	3(\rho\boldsymbol{\sigma})_{ij}\!\cdot\!\hat{\mathbf R}\,
	(\rho\boldsymbol{\sigma})_{kl}\!\cdot\!\hat{\mathbf R}
	\Big]~, \label{dotrho1}
\end{align}
where
 \begin{equation}
	\Gamma_{SS}(R)
	=
	\frac{\alpha}{4\pi m_f^2 R^3}
	\left[
	\operatorname{erf}\!\left(\frac{R}{2\sigma_0}\right)
	-
	\frac{2\sigma_0}{\sqrt{\pi}R}
	e^{-R^2/(4\sigma_0^2)}
	\right]~,
	\label{GammaSS10}
\end{equation}
 and $\sigma^\alpha$ ($\alpha=x,y,z$) are the Pauli matrices and
$\hat{\mathbf R}=\mathbf R/R$ is the unit vector along the inter-spin separation. One can verify by direct evaluation that the evolution equation
\eqref{dotrho1} is generated by a unitary commutator with the effective
spin-spin Hamiltonian
 The effective spin-spin interaction between the two qubits is described by the Hamiltonian
 \begin{equation}
 	H_{SS} = \Gamma_{SS}(R)\,
 	D_{\alpha\beta}
 	\,\sigma_A^\alpha \otimes \sigma_B^\beta ~,
 \end{equation}
 where the tensor
 \begin{equation}
 	D_{\alpha\beta} \equiv \delta_{\alpha\beta}-3\hat R_\alpha \hat R_\beta
 \end{equation}
 encodes the anisotropic structure of the dipole-dipole coupling.  
 The quantization axis is selected along the inter-spin direction,
 \(\hat{\mathbf R}=\hat z\). In this case, the nonvanishing components of
 \(D_{\alpha\beta}\) are
 \begin{align}
 	D_{xx} = 1~, \qquad
 	D_{yy} = 1~, \qquad
 	D_{zz} = -2~,
 \end{align}
 while all off diagonal entries vanish. The Hamiltonian therefore reduces
 to an anisotropic XXZ-type interaction
 \begin{equation}
 	H_{SS}
 	=
 	\Gamma_{SS}(R)
 	\left(
 	\sigma_x^A\sigma_x^B
 	+
 	\sigma_y^A\sigma_y^B
 	-
 	2\sigma_z^A\sigma_z^B
 	\right)~,
 	\label{eq:Hss-XXZ}
 \end{equation}
 and the density matrix of the two spin system evolves according to the von Neumann equation
 \begin{equation}
 	\frac{d\rho}{dt}=-i[H_{SS},\rho]~.
 \end{equation}
 We work in the standard product basis
 \begin{equation}
 	|1\rangle=|\uparrow\uparrow\rangle~,\quad
 	|2\rangle=|\uparrow\downarrow\rangle~,\quad
 	|3\rangle=|\downarrow\uparrow\rangle~,\quad
 	|4\rangle=|\downarrow\downarrow\rangle~,
 \end{equation}
 where $|ij\rangle = |i\rangle_A \otimes |j\rangle_B$, and define the matrix
 elements of the density operator as
 \begin{equation}
 	\rho_{IJ} \equiv \langle I|\rho|J\rangle~ ,
 	\qquad I,J=1,\ldots,4~ .
 \end{equation} 
 In this basis, the Hamiltonian \eqref{eq:Hss-XXZ} takes the matrix form
 \begin{equation}
 	H_{SS}
 	=
 	\Gamma
 	\begin{pmatrix}
 		-2 & 0 & 0 & 0 \\
 		0 & 2 & 2 & 0 \\
 		0 & 2 & 2 & 0 \\
 		0 & 0 & 0 & -2
 	\end{pmatrix},
 	\qquad
 	\Gamma \equiv \Gamma_{SS}(R)~.
 \end{equation}
 The states $|1\rangle$ and $|4\rangle$ are eigenstates of $H_{SS}$ with
 eigenvalue $-2\Gamma$, whereas the states $|2\rangle$ and $|3\rangle$ are
 coupled by the interaction. As a result, the evolution of the fully
 polarized states $|1\rangle$ and $|4\rangle$ amounts only to phase
 rotations under time evolution, while dynamics takes place
 within the $\{|2\rangle,|3\rangle\}$ subspace.
  For a generic initial density matrix
 \begin{equation}
 	\rho(0)=
 	\begin{pmatrix}
 		\rho_{11} & \rho_{12} & \rho_{13} & \rho_{14} \\
 		\rho_{21} & \rho_{22} & \rho_{23} & \rho_{24} \\
 		\rho_{31} & \rho_{32} & \rho_{33} & \rho_{34} \\
 		\rho_{41} & \rho_{42} & \rho_{43} & \rho_{44}
 	\end{pmatrix}~,
 \end{equation}
 the exact time evolution generated by $H_{SS}$ can be written as
 \begin{equation}
 	\rho(t)=
 	\begin{pmatrix}
 		\rho_{11} &
 		\rho_{12} e^{2 i\Gamma t} &
 		\rho_{13} e^{2 i\Gamma t} &
 		\rho_{14} \\
 		\rho_{21} e^{-2 i\Gamma t} &
 		\rho_{22}(t) &
 		\rho_{23}(t) &
 		\rho_{24} e^{-2 i\Gamma t} \\
 		\rho_{31} e^{-2 i\Gamma t} &
 		\rho_{32}(t) &
 		\rho_{33}(t) &
 		\rho_{34} e^{-2 i\Gamma t} \\
 		\rho_{41} &
 		\rho_{42} e^{2 i\Gamma t} &
 		\rho_{43} e^{2 i\Gamma t} &
 		\rho_{44}
 	\end{pmatrix}~.
 \end{equation}
 Here, the populations and coherences within the
 $\{|2\rangle,|3\rangle\}$ subspace evolve nontrivially. Introducing
 \[
 \rho_{22} \equiv \rho_{22}(0)~, \qquad
 \rho_{33} \equiv \rho_{33}(0)~, \qquad
 \rho_{23} \equiv \rho_{23}(0)~,
 \]
 with $\rho_{32}=\rho_{23}^*$, one obtains
 \begin{align}
 	\rho_{22}(t) &=
 	\rho_{22}\cos^2(2\Gamma t)
 	+
 	\rho_{33}\sin^2(2\Gamma t)
 	+
 	\Re(\rho_{23})\,\sin(4\Gamma t)~,
 	\label{eq:rho22-correct}
 	\\[2mm]
 	\rho_{33}(t) &=
 	\rho_{33}\cos^2(2\Gamma t)
 	+
 	\rho_{22}\sin^2(2\Gamma t)
 	-
 	\Re(\rho_{23})\,\sin(4\Gamma t)~,
 	\label{eq:rho33-correct}
 	\\[2mm]
 	\rho_{23}(t) &=
 	\rho_{23}\cos^2(2\Gamma t)
 	+
 	\rho_{32}\sin^2(2\Gamma t)
 	+
 	\frac{i}{2}\bigl(\rho_{22}-\rho_{33}\bigr)\sin(4\Gamma t)~,
 	\label{eq:rho23-correct}
 \end{align}
 together with
 \begin{equation}
 	\rho_{32}(t)=\rho_{23}^*(t)~.
 \end{equation}
 Equivalently, using $\rho_{32}=\rho_{23}^*$, the coherence may be written as
 \begin{equation}
 	\rho_{23}(t)
 	=
 	\Re(\rho_{23})
 	+
 	i\,\Im(\rho_{23})\cos(4\Gamma t)
 	+
 	\frac{i}{2}
 	\bigl(\rho_{22}-\rho_{33}\bigr)\sin(4\Gamma t)~.
 \end{equation}
  One can verify that
 \begin{equation}
 	\rho_{22}(t)+\rho_{33}(t)
 	=
 	\rho_{22}+\rho_{33}~,
 \end{equation}
 remains constant.
  Equations \eqref{eq:rho22-correct}--\eqref{eq:rho23-correct} make explicit how the dipole induced spin-spin interaction generates coherent population transfer between the states
 $|\uparrow\downarrow\rangle$ and $|\downarrow\uparrow\rangle$, while coherences involving the fully polarized states
 $|\uparrow\uparrow\rangle$ and $|\downarrow\downarrow\rangle$ evolve only through phase rotations. This structure reflects the conservation of the total spin projection $S_z$ and the block diagonal form of $H_{SS}$ in the chosen basis.
  
  	\begin{figure}[t]
  	\centering
  	\vspace{-1cm}
  	\includegraphics[scale=0.3]{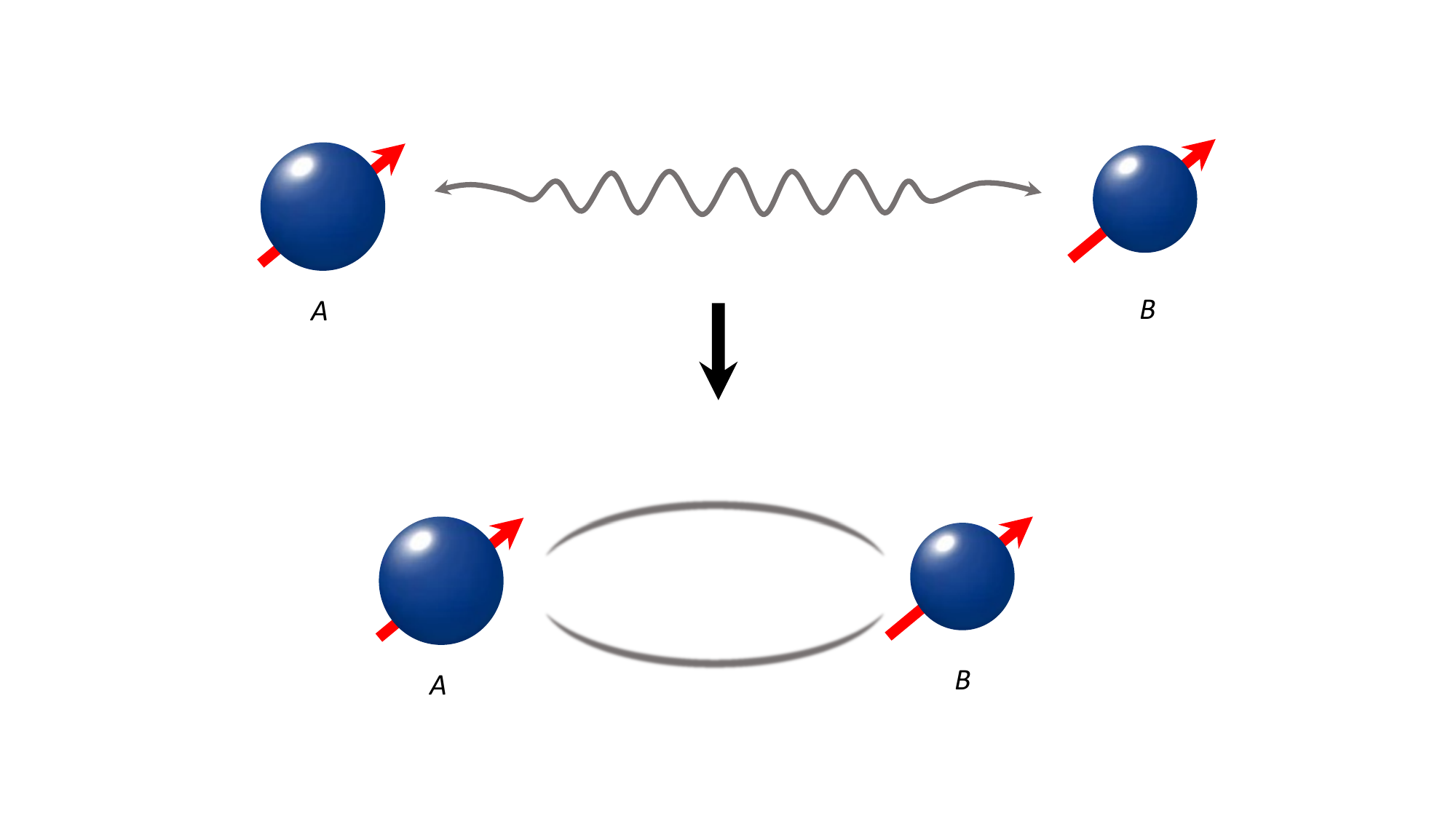}
  	\vspace{-1cm}
  	\caption{  The illustration of photon mediated spin-spin interaction between two localized fermionic qubits. Spin 
  		$A$
  		and spin 
  		$B$
  		are separated by a distance 
  		$R$
  		. The wavy line represents the exchange of a virtual photon that mediates an effective interaction between the spins.}
  	\label{fig1}
  \end{figure}

   \subsection{Negativity and entanglement generation}
   
   For a two qubit density matrix $\rho(t)$, the entanglement negativity is defined as \cite{ Horodecki:2009zz}
   \begin{equation}
   	\mathcal{N}(t)
   	=
   	\frac{1}{2}\bigl(\|\rho^{T_B}(t)\|_1 - 1\bigr)
   	=
   	\sum_{\lambda_i<0} |\lambda_i|~,
   	\label{Nt}
   \end{equation}
   where $\rho^{T_B}$ denotes the partial transpose of $\rho$ with respect to
   subsystem $B$, $\|\cdot\|_1$ is the trace norm, and $\lambda_i$ are the
   eigenvalues of the partially transposed density matrix $\rho^{T_B}$.
   Entanglement is present if and only if
  $
   	\mathcal{N}(t) > 0
$.
   The effective spin--spin interaction is governed by the Hamiltonian
   $H_{SS}$, which satisfies the commutation relation
   \begin{equation}
   	[H_{SS},S_z]=0~,
   \end{equation}
   where $S_z$ is the total spin projection operator. This symmetry implies
   that the dynamics preserves the total spin projection and therefore
   decomposes the Hilbert space into invariant sectors
   \begin{equation}
   	\{|\uparrow\uparrow\rangle\}~,
   	\qquad
   	\{|\uparrow\downarrow\rangle,|\downarrow\uparrow\rangle\}~,
   	\qquad
   	\{|\downarrow\downarrow\rangle\}~.
   \end{equation}
    To isolate the entanglement generated solely by the spin-spin
   interaction, we consider an initially separable state
   \begin{equation}
   	\rho(0)=\rho^A\otimes\rho^B~,
   \end{equation}
   which is further assumed to be diagonal in the computational basis
 $
   \{
   |\uparrow\uparrow\rangle,
   |\uparrow\downarrow\rangle,
   |\downarrow\uparrow\rangle,
   |\downarrow\downarrow\rangle
   \}.
  $
   Because $H_{SS}$ only couples states within the same $S_z$ sector,
   the evolution mixes exclusively the states
   $|\uparrow\downarrow\rangle$ and $|\downarrow\uparrow\rangle$.
   Consequently, if coherences between different $S_z$ sectors are absent
   initially, they remain zero throughout the evolution. Under these
   conditions, the reduced two spin density matrix retains the X-state structure
   \begin{equation}
   	\rho(t)=
   	\begin{pmatrix}
   		\rho_{11}(t) & 0 & 0 & \rho_{14}(t) \\
   		0 & \rho_{22}(t) & \rho_{23}(t) & 0 \\
   		0 & \rho_{32}(t) & \rho_{33}(t) & 0 \\
   		\rho_{41}(t) & 0 & 0 & \rho_{44}(t)
   	\end{pmatrix}~.
   \end{equation}
   The matrix decomposes into two independent \(2\times2\) blocks,
   corresponding to the subspaces
$
   \{|1\rangle,|4\rangle\}~,
   \qquad
   \{|2\rangle,|3\rangle\}.
$
    After computing the eigenvalues of the corresponding partial transpose matrix with respect to qubit \(B\) one finds the negativity as
   \begin{equation}
   	\begin{aligned}
   		\mathcal{N}(t)
   		&=
   		\max\!\left[
   		0,
   		\frac{
   			\sqrt{
   				\bigl(\rho_{11}(t)-\rho_{44}(t)\bigr)^2
   				+
   				4|\rho_{23}(t)|^2
   			}
   			-
   			\bigl(\rho_{11}(t)+\rho_{44}(t)\bigr)
   		}{2}
   		\right]
   		\\[2mm]
   		&\quad+
   		\max\!\left[
   		0,
   		\frac{
   			\sqrt{
   				\bigl(\rho_{22}(t)-\rho_{33}(t)\bigr)^2
   				+
   				4|\rho_{14}(t)|^2
   			}
   			-
   			\bigl(\rho_{22}(t)+\rho_{33}(t)\bigr)
   		}{2}
   		\right]~.
   	\end{aligned}
   \end{equation}
     For an initially separable pure state one has $
   	\mathcal{N}(0)=0
$. The interaction Hamiltonian \(H_{SS}\) dynamically generates the
   coherence \(\rho_{23}(t)\) within the
   \(\{|2\rangle,|3\rangle\}\) subspace.
   For an initially diagonal state,
   \begin{equation}
   	\rho_{23}(0)=0,
   \end{equation}
   and the general evolution derived previously reduces to
   \begin{equation}
   	\rho_{23}(t)
   	=
   	\frac{i}{2}
   	\bigl[\rho_{22}(0)-\rho_{33}(0)\bigr]
   	\sin\!\bigl(4\Gamma_{SS}(R)t\bigr)~.
   \end{equation}
    For the initial state
  $
   	\rho(0)
   	=
   	|\uparrow\downarrow\rangle
   	\langle\uparrow\downarrow|
$ one has
$
   	\rho_{22}(0)=1$ and $
   	\rho_{33}(0)=0
$
   that yields
   \begin{equation}
   	\rho_{23}(t)
   	=
   	\frac{i}{2}
   	\sin\!\bigl(4\Gamma_{SS}(R)t\bigr)~,
   \end{equation}
   and therefore
   \begin{equation}
   	|\rho_{23}(t)|
   	=
   	\frac12
   	\left|
   	\sin\!\bigl(4\Gamma_{SS}(R)t\bigr)
   	\right|~.
   \end{equation}
      Since in this case
 $
   \rho_{11}(t)=\rho_{44}(t)=0$ and 
   $
   \rho_{14}(t)=0
   $
   the only potentially negative eigenvalue is
   \begin{equation}
   	\lambda_-
   	=
   	-
   	|\rho_{23}(t)|~.
   \end{equation}
   Consequently, the negativity becomes
   \begin{equation}
   	\mathcal{N}(t)
   	=
   	\frac12
   	\left|
   	\sin\!\bigl(4\Gamma_{SS}(R)t\bigr)
   	\right|~.
   \end{equation}
   The negativity reaches its maximum value
   \begin{equation}
   	\mathcal{N}_{\max}=\frac12~,
   \end{equation}
   at
   \begin{equation}
   	t=\frac{\pi}{8\Gamma_{SS}(R)}~.
   \end{equation}
   At this time the system occupies a maximally entangled state within the
   single excitation subspace, locally equivalent to a Bell state.

 The physical origin of the entanglement generation can be understood by relating the effective spin-spin interaction to the well-known XY exchange model. The dominant dynamical process in Eq.~\eqref{eq:Hss-XXZ} is governed by the flip-flop terms
 \begin{equation}
 	\sigma_x^A\sigma_x^B + \sigma_y^A\sigma_y^B~,
 \end{equation}
 which induce coherent exchange of single spin excitations between the states $|\uparrow\downarrow\rangle$ and $|\downarrow\uparrow\rangle$ while preserving the total excitation number.
  This structure is captured by the standard XY Hamiltonian
 \begin{equation}
 	H_{XY}
 	=
 	J \left(
 	\sigma_x^A \sigma_x^B
 	+
 	\sigma_y^A \sigma_y^B
 	\right)
 	=
 	2J \left(
 	\sigma_+^A \sigma_-^B
 	+
 	\sigma_-^A \sigma_+^B
 	\right)~,
 \end{equation}
 which generates Rabi oscillations in the single excitation subspace with frequency $2J$. The corresponding unitary evolution
 \begin{equation}
 	U(t)=e^{-iH_{XY}t}~,
 \end{equation}
 implements an iSWAP-type entangling gate. At the characteristic time
 \begin{equation}
 	t_{\mathrm{iSWAP}}=\frac{\pi}{4J}~,
 \end{equation}
 the system undergoes the transformation
 \begin{equation}
 	|\uparrow\downarrow\rangle \rightarrow i|\downarrow\uparrow\rangle~,
 	\qquad
 	|\downarrow\uparrow\rangle \rightarrow i|\uparrow\uparrow\rangle~,
 \end{equation}
 while leaving the fully polarized states unchanged. This evolution generates maximally entangled Bell states from initially separable configurations.
  In the present QED derived model, the effective Hamiltonian
 \begin{equation}
 	H_{SS}(R)
 	\propto
 	\sigma_x^A\sigma_x^B
 	+
 	\sigma_y^A\sigma_y^B
 	-
 	2\sigma_z^A\sigma_z^B~,
 \end{equation}
 contains precisely this XY exchange interaction as its dynamical core, supplemented by an additional Ising type contribution. The XY sector governs coherent excitation exchange and is responsible for the time dependent off diagonal coherence $\rho_{23}(t)$, while the $ZZ$ term produces an energy shift that modifies the oscillation frequency but does not qualitatively affect entanglement generation.

	\begin{figure}[t]
		\centering
		\vspace{-1cm}
		\includegraphics[scale=0.4]{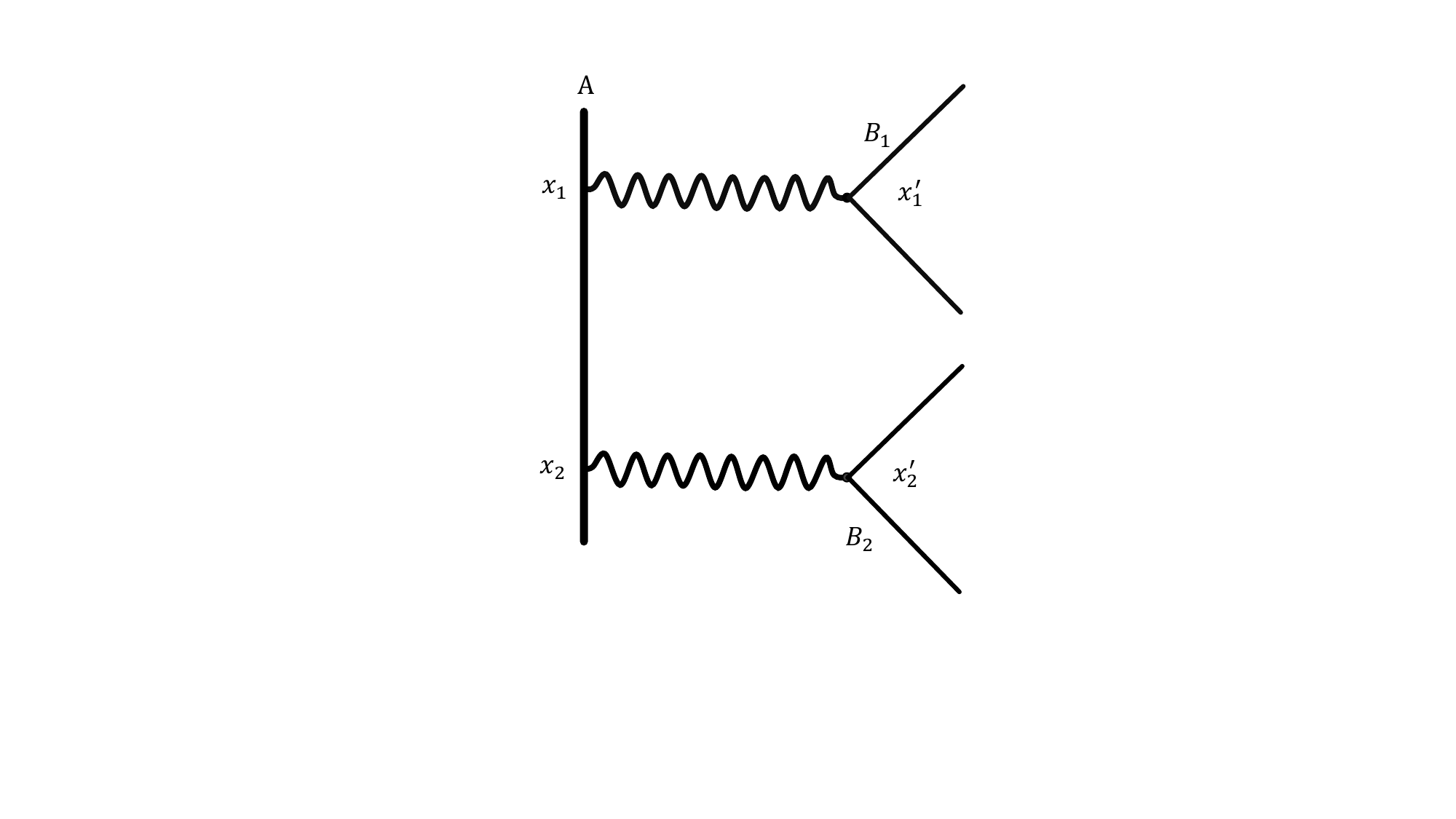}
		\vspace{-2cm}
		\caption{ The Feynmann diagram asociated with interavtion of qubit (fermion) $A$ with $2$ qubits $B_1$ and $B_2$.}
		\label{fig2}
	\end{figure} 
	
	\section{Interaction of qubit A with $N=2$ qubits }

In this section, we extend the two qubit analysis to a scenario in which qubit \(A\) interacts sequentially with two qubits, \(B_1\) and \(B_2\), through photon exchange. Specifically, fermion \(A\) propagates from \(x_1 \to x_2\), interacting first with \(B_1\) at \(x'_1\) and subsequently with \(B_2\) at \(x'_2\) (Fig. \eqref{fig2}).

The effective interaction describing this sequential photon mediated coupling can be written as
\begin{equation}
	\begin{aligned}
		\hat H_{\mathrm{eff}} 
		&= q^4 V^4 
		\int d\tau_1\, d\tau_2\, d\tau'_1\, d\tau'_2
		\int d^3 x_1\, d^4 x_2\, d^4 x'_1\, d^4 x'_2 
		\bar \psi^-_A(x_1)\, \gamma^{\mu_1}\, 
		D_F(x_1 - x_2)\, 
		\gamma^{\mu_2}\, 
		\psi^+_A(x_2)  \\
		&\quad \times 
		P_{F\,\mu_1 \nu_1}(x_1 - x'_1)\,
		\bar \psi^-_{B_1}(x'_1)\, \gamma^{\nu_1}\, \psi^+_{B_1}(x'_1) 
		P_{F\,\mu_2 \nu_2}(x_2 - x'_2)\,
		\bar \psi^-_{B_2}(x'_2)\, \gamma^{\nu_2}\, \psi^+_{B_2}(x'_2) \\
		&\quad \times 
		\delta^{(4)}_{\sigma_A}(x_1 - \bar x_{1A}(\tau_1))\,
		\delta^{(4)}_{\sigma_A}(x_2 - \bar x_{2A}(\tau_2)) 
		\delta^{(4)}_{\sigma_{1B}}(x'_1 - \bar x_{1B}(\tau'_1))\,
		\delta^{(4)}_{\sigma_{2B}}(x'_2 - \bar x_{2B}(\tau'_2))~ ,
	\end{aligned}
	\label{H93}
\end{equation}
where \(D_F(x_1-x_2)\) denotes the Feynman propagator describing the propagation of fermion \(A\) between the spacetime points \(x_1\) and \(x_2\). Its Fourier representation is
\begin{equation}
	D_F(x_1-x_2)
	=
	\int \frac{d^4Q}{(2\pi)^4}
	\frac{i(\slashed Q + m_A)}{Q^2 - m_A^2 + i\epsilon}
	e^{-iQ\cdot(x_1-x_2)}~ ,
\end{equation}
and the photon Feynman propagator mediating the interaction between qubit \(A\) and the qubits \(B_1\) and \(B_2\). The operators
\(\bar \psi_{B_1} \gamma^{\nu_1} \psi_{B_1}\) and
\(\bar \psi_{B_2} \gamma^{\nu_2} \psi_{B_2}\)
represent the local Dirac currents of the qubits \(B_1\) and \(B_2\).
Also, \(\delta^{(4)}_\sigma\) account for the finite spatial localization of the qubits, as introduced in the previous section.

		After performing the momentum integrations and retaining only the
		spin-spin contribution (see Appendix~B for the detailed derivation),
		the reduced evolution equation becomes
		\begin{equation}
			\dot{\rho}_{IJ}
			=
			i\frac{q^4}{4m_A m_{B_1}m_{B_2}}
			T_{\alpha\beta}\,
			\rho^A_{ij}
			\left[
			(\sigma^\alpha\rho^{B_1})_{k_1l_1}
			(\sigma^\beta\rho^{B_2})_{k_2l_2}
			-
			(\rho^{B_1}\sigma^\alpha)_{k_1l_1}
			(\rho^{B_2}\sigma^\beta)_{k_2l_2}
			\right],
		\end{equation}
		where the composite indices are defined by
		\[
		I=(i,k_1,k_2),
		\qquad
		J=(j,l_1,l_2).
		\]
		
		The spatial tensor entering the interaction kernel is
		\begin{equation}
			T_{\alpha\beta}
			=
			\delta_{\alpha\beta}\,
			\nabla_{\mathbf r_{B_1}}F_{a_1}
			\!\cdot\!
			\nabla_{\mathbf r_{B_2}}F_{a_2}
			-
			(\nabla_{\mathbf r_{B_2}}F_{a_2})_\alpha
			(\nabla_{\mathbf r_{B_1}}F_{a_1})_\beta~,
			\label{T12}
		\end{equation}
		where the spatial kernel is given by the regularized Coulomb potential
		\begin{equation}
			F_{a_i}(r)
			=
			\frac{1}{4\pi r}\,
			\mathrm{erf}\!\left(
			\frac{r}{2\sqrt{a_i}}
			\right).
		\end{equation}
		
		Its gradient is purely radial and takes the form
		\begin{equation}
			\nabla F_{a_i}(r)
			=
			-
			\frac{\hat{\mathbf r}}{4\pi r^2}
			\left[
			\mathrm{erf}\!\left(
			\frac{r}{2\sqrt{a_i}}
			\right)
			-
			\frac{r}{\sqrt{\pi a_i}}
			e^{-r^2/(4a_i)}
			\right].
		\end{equation}
		
		The evolution equation therefore acquires the structure of an effective
		spin-spin interaction, where the Pauli matrices generated by the
		spinor bilinears act on the bath spin density matrices, while the
		spatial dependence of the interaction is encoded entirely in the tensor
		\(T_{\alpha\beta}\).

	 \subsection{Dipole-dipole tensor for $B_1$ and $B_2$ qubits}
	 
	 We now turn to evaluate the effective interaction generated by the
	 three body QED process. Starting from the coordinate space
	 tensor in Eq.~\eqref{T12}, the bath separation vector is defined as
	 \begin{equation}
	 	\mathbf R
	 	=
	 	\mathbf r_{B_1}-\mathbf r_{B_2},
	 	\qquad
	 	R=|\mathbf R|,
	 	\qquad
	 	\hat{\mathbf R}=\frac{\mathbf R}{R}~.
	 \end{equation} 
	 For separations $R\gg \sqrt{a_i}$, the regularized Coulomb kernel reduces to
	 \begin{equation}
	 	\nabla F_{a_i}(R)
	 	\simeq
	 	-\frac{\hat{\mathbf R}}{4\pi R^2}~.
	 \end{equation}
	 Hence,
	 \begin{equation}
	 	(\nabla_{\mathbf r_{B_1}}F_{a_1})_\alpha
	 	\simeq
	 	-\frac{\hat R_\alpha}{4\pi R^2},
	 	\qquad
	 	(\nabla_{\mathbf r_{B_2}}F_{a_2})_\beta
	 	\simeq
	 	+\frac{\hat R_\beta}{4\pi R^2}~,
	 \end{equation}
	 which, when substituted into Eq.~\eqref{T12}, gives
	 \begin{equation}
	 	T_{\alpha\beta}
	 	\simeq
	 	\frac{1}{(4\pi)^2 R^4}
	 	\left(
	 	\hat R_\alpha \hat R_\beta
	 	-
	 	\delta_{\alpha\beta}
	 	\right)~.
	 	\label{Tab-final}
	 \end{equation}
	 The interaction generated by sequential virtual photon exchange is therefore
	 \begin{equation}
	 	H_{AB_1B_2}
	 	=
	 	\frac{q^4}{4m_A m_{B_1}m_{B_2}}\,
	 	T_{\alpha\beta}\,
	 	\mathbb{1}_A \otimes
	 	\sigma_{B_1}^{\alpha}\otimes
	 	\sigma_{B_2}^{\beta}~.
	 	\label{HAB12}
	 \end{equation}
	 	 The identity operator $\mathbb{1}_A$ reflects that, at leading
	 nonrelativistic order, the mediator spin does not participate in the
	 interaction. This follows from the reduction of the mediator current,
	 which yields the trivial spin contraction
$
	 	\chi^\dagger_{r_A'}\chi_{r_A}
	 	=
	 	\delta_{r_A r_A'}
$
 so that no Pauli matrices associated with particle $A$ appear in the kernel. 
	 Tracing over the mediator spin
	 \begin{equation}
	 	\mathrm{Tr}_A\!\left(\mathbb{1}_A\rho^A\right)
	 	=
	 	\mathrm{Tr}(\rho^A)=1~,
	 \end{equation}
	 the effective Hamiltonian acting on the bath subsystem reduces to
	 \begin{align}
	 	H_{B_1B_2}
	 	=
	 	J(R)\left[
	 	(\boldsymbol{\sigma}_{B_1}\!\cdot\!\hat{\mathbf R})
	 	(\boldsymbol{\sigma}_{B_2}\!\cdot\!\hat{\mathbf R})
	 	-
	 	\boldsymbol{\sigma}_{B_1}\!\cdot\!\boldsymbol{\sigma}_{B_2}
	 	\right]~, \label{HB12-final}
	 \end{align}
	 with
	 \begin{equation}
	 	J(R)=
	 	\frac{q^4}{64\pi^2 m_A m_{B_1}m_{B_2}}
	 	\frac{1}{R^4}~.
	 	\label{JR-final}
	 \end{equation}
	  The interaction is anisotropic and decays as $R^{-4}$. The mediator $A$
	 acts only virtually, setting the overall scale $J(R)$.

	 \begin{figure}[t]
	 	\centering
	 	\vspace{-1cm}
	 	\includegraphics[scale=0.3]{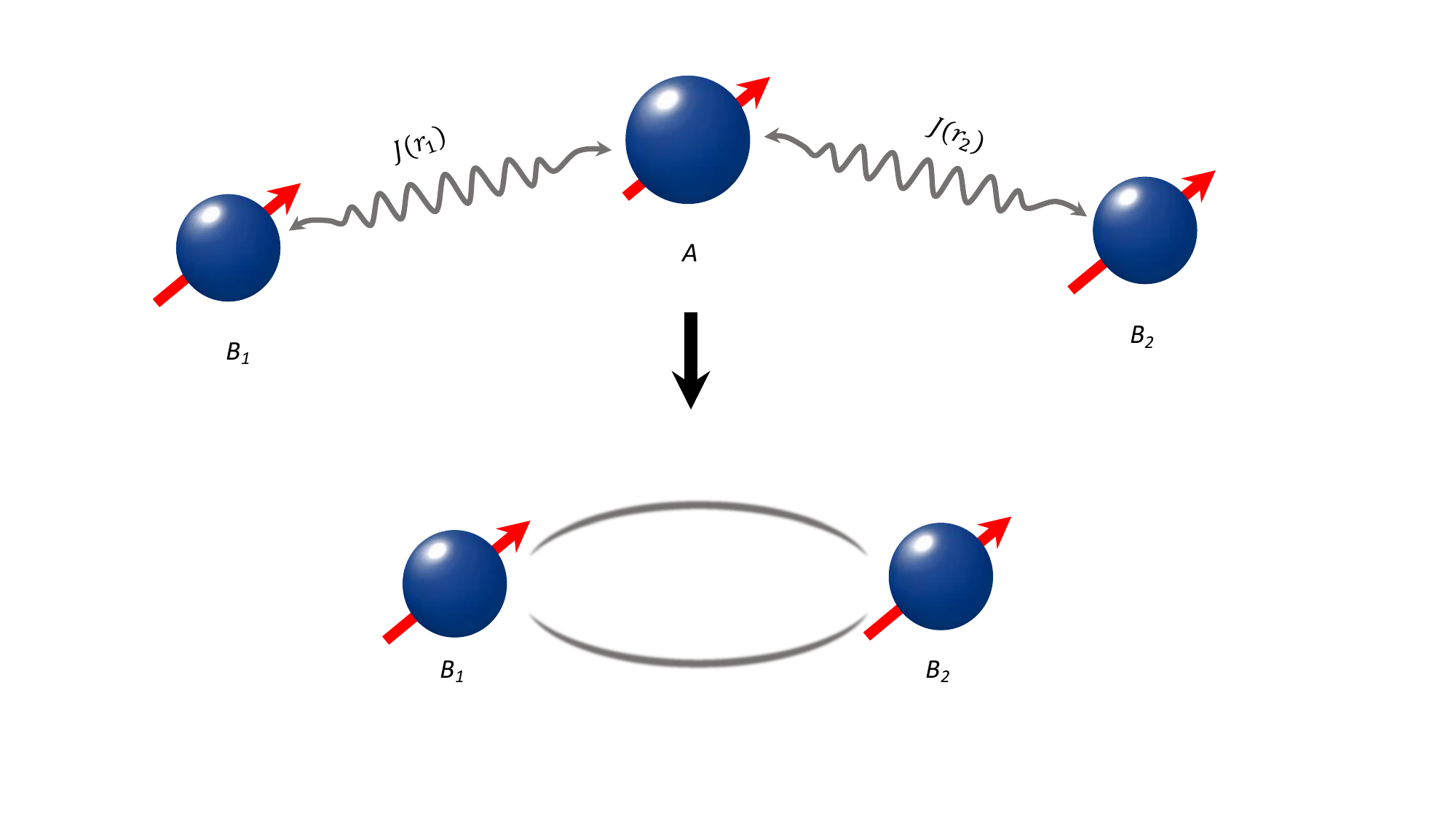}
	 	\vspace{-1cm}
	 	\caption{ The illustration of photon mediated spin-spin interaction between three localized fermionic qubits. }
	 	\label{fig3}
	 \end{figure}

\subsubsection{Two qubit reduced dynamics and entanglement genration }

At leading nonrelativistic order, the mediator spin appears only through
the identity operator and therefore does not affect the dynamical evolution
of the bath subsystem. 
For an initially factorized state,
\begin{equation}
	\rho_{\mathrm{tot}}(0)=\rho_A\otimes\rho_{B_1B_2}(0)~,
\end{equation}
the unitary evolution preserves this product structure,
\begin{equation}
	\rho_{\mathrm{tot}}(t)=\rho_A\otimes\rho_{B_1B_2}(t)~,
\end{equation}
and the reduced state is
\begin{equation}
	\rho(t)\equiv\rho_{B_1B_2}(t)
	=\mathrm{Tr}_A[\rho_{\mathrm{tot}}(t)]~.
\end{equation}
Choosing the quantization axis along the separation vector
$\hat{\mathbf z}=\hat{\mathbf R}$, the Hamiltonian
Eq.~\eqref{HB12-final} becomes
\begin{align}
	H_{\mathrm{eff}}
	=
	-J(R)\left(
	\sigma_{B_1}^x\sigma_{B_2}^x+
	\sigma_{B_1}^y\sigma_{B_2}^y
	\right)
	=
	-2J(R)\left(
	\sigma_{B_1}^+\sigma_{B_2}^-+
	\sigma_{B_1}^-\sigma_{B_2}^+
	\right)~,
	\label{Heff}
\end{align}
which generates coherent flip-flop dynamics in the
single excitation subspace
$\{|\uparrow\downarrow\rangle,|\downarrow\uparrow\rangle\}$.
In the computational basis
\begin{equation}
	\{|\uparrow\uparrow\rangle,
	|\uparrow\downarrow\rangle,
	|\downarrow\uparrow\rangle,
	|\downarrow\downarrow\rangle\}~,
\end{equation}
the state evolves unitarily as
\begin{equation}
	\rho(t)=e^{-iH_{\mathrm{eff}}t}\rho(0)e^{iH_{\mathrm{eff}}t}~.
\end{equation}
The dynamics is restricted to the single excitation sector with frequency
\begin{equation}
	\Omega = 4J(R)~.
\end{equation}
For the initial state $|\psi(0)\rangle=|\uparrow\downarrow\rangle$
\begin{equation}
	|\psi(t)\rangle
	=
	\cos\!\left(\frac{\Omega t}{2}\right)|\uparrow\downarrow\rangle
	-
	i\sin\!\left(\frac{\Omega t}{2}\right)|\downarrow\uparrow\rangle~.
\end{equation}
The corresponding density matrix elements yield the entanglement negativity
\begin{equation}
	\mathcal N(t)=\frac{1}{2}|\sin(\Omega t)|~.
\end{equation}
Maximal entanglement $\mathcal N_{\max}=1/2$ is reached at
\begin{equation}
	t_{\max}=\frac{\pi}{2\Omega}=\frac{\pi}{8J(R)}~,
\end{equation}
corresponding to the Bell state
\begin{equation}
	|\psi\rangle=
	\frac{1}{\sqrt{2}}\left(
	|\uparrow\downarrow\rangle
	-i|\downarrow\uparrow\rangle
	\right)~,
\end{equation}
and the entanglement generation time scales as
\begin{equation}
	t_{\max}\propto m_A m_{B_1}m_{B_2}R^4~.
\end{equation}
The mediator $A$ thus acts as a virtual channel that fixes the coupling
scale $J(R)$, while all entanglement dynamics occurs entirely within the
bath subsystem (Fig. \eqref{fig3}).

  \section{Extension to a bath of $N$ qubits}
  
  We now generalize the previous calculations to a system in which a single
  mediator qubit $A$ interacts with a collection of bath qubits
  $\{B_1,B_2,\dots,B_N\}$. Each bath qubit couples locally to the mediator
  via the same QED vertex considered in the previous sections.
  The total Hilbert space is
  \begin{equation}
  	\mathcal{H}
  	=
  	\mathcal{H}_A
  	\otimes
  	\bigotimes_{i=1}^{N}\mathcal{H}_{B_i}~,
  \end{equation}
  and the total density matrix has dimension $2^{N+1}$.
  
 At the microscopic level, the interaction corresponds to a sequence of
  mediator–bath scattering events. For $N$ bath qubits, the perturbative
  order increases as $q^{2N}$, reflecting $N$ photon exchanges between the
  mediator and the bath degrees of freedom. Schematically, the interaction
  amplitude can be written as
  \begin{align}
  	H_{\mathrm{int}}^{(N)}(t)
  	\;\propto\;
  	q^{2N}
  	\!\!\!\sum_{\{r\}}
  	\int dQ
  	\prod_{j=1}^{N}
  	\Big[
  	dK_j\, d^4x_j\, d^4x_j'\, d\tau_j'\;
  	u^\dagger_{B_j}(k_j')
  	\gamma^{\nu_j}
  	u_{B_j}(k_j)\,
  	D_{\mu_j\nu_j}(K_j)\,
  	\delta^{(4)}(x_j'-\bar{x}_{jB}(\tau_j'))
  	\Big]
  	\nonumber\\
  	\times
  	u^\dagger_A(p_1)\gamma^{\mu_1}
  	\frac{i(\slashed{Q}+m_A)}{Q^2-m_A^2+i\epsilon}
  	\gamma^{\mu_N}u_A(p_2)\,
  	e^{-i(p_1-Q-K_1)\cdot x_1}
  	e^{-i(Q-p_2-K_N)\cdot x_N}~.
  	\label{eq:H_N_bath}
  \end{align}
  
  Each bath vertex contributes a Dirac bilinear, a photon propagator, and a
  spatial smearing kernel. The mediator carries a single internal momentum
  $Q$, connecting all interaction segments.
  At this level the amplitude scales as
  \begin{equation}
  	\mathcal{M}^{(N)}
  	\sim
  	q^{2N}
  	\left(m_A \prod_{j=1}^{N} m_{B_j}\right)^{-1}
  	\times (\text{propagators and spatial kernels})~,
  	\label{eq:M_N_scaling}
  \end{equation}
  up to geometrical and regularization dependent prefactors.
 Despite the apparent $N$-body structure of Eq.~\eqref{eq:H_N_bath},
  the reduced dynamics of the bath is dominated by pairwise contractions.
  After tracing over the mediator and photon degrees of freedom and applying
  the Born-Markov expansion, only two point correlation functions survive,
  since the photon propagator is Gaussian and supports no higher order
  connected contractions. As a result, the effective kernel factorizes as
  \begin{equation}
  	\mathcal{K}_{ij}
  	\propto
  	\langle J_A J_A\rangle
  	\langle J_{B_i} J_{B_j}\rangle~,
  	\qquad i\neq j~,
  	\label{eq:pairwise_kernel}
  \end{equation}
  and higher order bath cumulants do not contribute at leading order.
  Each pair $(i,j)$ is associated with the same geometric tensor structure
  encountered in the two- and three-qubit cases,
  \begin{equation}
  	T_{\alpha\beta}^{(ij)}
  	=
  	\delta_{\alpha\beta}\,
  	\nabla_i F_{a_i}(\mathbf r_i)\cdot\nabla_j F_{a_j}(\mathbf r_j)
  	-
  	(\nabla_j F_{a_j})_\alpha
  	(\nabla_i F_{a_i})_\beta~,
  	\label{eq:Tij_N}
  \end{equation}
  where $F_a(r)$ is the regularized Coulomb kernel.
  For large separations $R_{ij}\gg \sqrt{a_i,a_j}$,
  \begin{equation}
  	\nabla F_{a_i}(R_{ij})
  	\simeq
  	-\frac{\hat{\mathbf R}_{ij}}{4\pi R_{ij}^2}~,
  \end{equation}
  which leads to the asymptotic form
  \begin{equation}
  	T_{\alpha\beta}^{(ij)}
  	=
  	\frac{1}{(4\pi)^2 R_{ij}^4}
  	\left(
  	\hat R_{ij,\alpha}\hat R_{ij,\beta}
  	-
  	\delta_{\alpha\beta}
  	\right)~.
  	\label{eq:Tij_asymp}
  \end{equation}
   After contraction with the Pauli operators, the bath dynamics reduces to a
  fully connected XY spin network
  \begin{equation}
  	H_{\mathrm{eff}}^{(N)}
  	=
  	\sum_{1\le i<j\le N}
  	J_{ij}
  	\left(
  	\sigma^x_{B_i}\sigma^x_{B_j}
  	+
  	\sigma^y_{B_i}\sigma^y_{B_j}
  	\right)~,
  	\label{eq:Heff_N}
  \end{equation}
  where the exchange couplings are
  \begin{equation}
  	J_{ij}
  	=
  	\frac{q^4}{64\pi^2 m_A m_{B_i} m_{B_j}}
  	\frac{C_{ij}}{R_{ij}^4}~.
  	\label{eq:Jij}
  \end{equation}
 Here $C_{ij}=O(1)$ encodes orientation dependent factors arising from
  \eqref{eq:Tij_asymp}.
The interaction is pairwise and long range, generating a fully
  connected anisotropic XY model on the bath subsystem. The mediator $A$
  does not appear as an active spin degree of freedom in the reduced
  dynamics; instead, it sets the overall interaction scale through the
  factor $1/m_A$ in Eq.~\eqref{eq:Jij}. Higher order multi bath
  correlations are suppressed at leading order due to the Gaussian structure
  of the photon propagator and the factorization of connected diagrams.

 \section{Conclusion}

  We have presented a microscopic derivation of entanglement generation between spatially separated spin-$1/2$ systems generated by QED interactions. Starting from the underlying QED interaction Hamiltonian, we systematically integrated out the photon field and the intermediate fermionic degrees of freedom, obtaining an effective description of the reduced spin dynamics in terms of an spin-spin Hamiltonian.
  
  The effective interactions are determined by the tensor structure of the exchanged photon propagators, which generate anisotropic spin-dependent couplings between the fermionic spins. For the direct mediator--spin interaction (\(A\!-\!B\)), the nonrelativistic reduction reproduces the standard dipolar spin--spin structure with the familiar long-distance behavior proportional to \(R^{-3}\). In contrast, the induced interaction between the bath spins (\(B_1\!-\!A\!-\!B_2\)), generated through the sequential exchange process mediated by the intermediate fermion \(A\), acquires an additional spatial suppression. As a result, the corresponding effective coupling exhibits a \(R^{-4}\) decaying behaviour. 
  
  Within the perturbative regime considered here, the mediator particle does not appear as a dynamical degree of freedom in the effective theory. Instead, it contributes only virtually, entering the reduced description through effective coupling constants that depend on the particle masses and the electromagnetic coupling. 
  In this way, the effective Hamiltonian encodes the underlying QED scattering process.
  
  The analysis of the two-spin dynamics shows that coherent exchange in the single-excitation sector generates entanglement between spatially separated subsystems. Using the entanglement negativity as a measure, we demonstrate that nonclassical correlations arise dynamically from initially separable states and are fully driven by the effective interaction. The strength and time dependence of entanglement can be controlled via both the interspin separation and the geometric configuration of the system.
  
  The present formulation 
  relates QED scattering amplitudes to effective spin couplings in the reduced dynamics.
  By mapping field theoretic propagator structures onto effective spin Hamiltonians, we clarify how relativistic field mediated interactions generate entanglement between distant quantum systems. 
  The construction can be extended to larger spin networks generated by repeated mediator exchange.

	\section{Acknowledgements}
	
	MZ would like to thank M. Abadi for very useful discussions and comments.
	
	\bibliography{refs.bib}

	\appendix
	
	\section{Details of the derivation of the master equation for $A$-$B$ interaction}
	Integrating out the electromagnetic field to second order yields an effective interaction between the localized fermionic systems. Starting from the QED interaction Hamiltonian and eliminating the photon field, the effective Hamiltonian can be written as
	\begin{equation}
		\hat H_{\mathrm{eff}}
		= VV' \!\int d\tau\, d\tau' \!\int d^{3}x\, d^{4}x'\;
		\big[\bar\psi(x)\gamma^\mu\psi(x)\big]\,
		iP_{F\,\mu\nu}(x-x')\,
		\big[\bar\psi(x')\gamma^\nu\psi(x')\big]\,
		\delta_{\sigma_0}^{4}\!\big(x-\bar x(\tau)\big)\,
		\delta_{\sigma_0'}^{4}\!\big(x'-\bar x'(\tau')\big),
	\end{equation}
	where \(P_{F\,\mu\nu}(x-x')\) denotes the Feynman propagator of the electromagnetic field and \(\bar x(\tau)\) and \(\bar x'(\tau')\) describe the spacetime trajectories of the two localized systems. The functions \(\delta_{\sigma_0}^{4}\) and \(\delta_{\sigma_0'}^{4}\) characterize the finite spacetime smearing of the interaction region around each trajectory.
	To evaluate the fermionic currents we decompose the Dirac field into positive- and negative-frequency components
	\begin{equation}
		\psi(x)=\psi^{+}(x)+\psi^{-}(x)~,
	\end{equation}
	where
	\begin{eqnarray}
		\psi^{+}(x)
		&=&
		\int \frac{d^{3}p}{(2\pi)^{3}}
		\sum_{r}
		u_{r}(\mathbf p)\,
		\hat c_{r}(\mathbf p)\,
		e^{-ip\cdot x}~,\\
		\psi^{-}(x) 
		&=&
		\int \frac{d^{3}p}{(2\pi)^{3}}
		\sum_{r}
		u_{r}^{\dagger}(\mathbf p)\,
		\hat c_{r}^{\dagger}(\mathbf p)\,
		e^{ip\cdot x}~,
\end{eqnarray}
	in which the fermionic creation and annihilation operators satisfy the  following canonical anticommutation relations
	\begin{align}
	\left\{\, \hat{c}_r({\bf p})\,,\,\hat{c}^\dag_{r'}({\bf p'})\,\right\}
	&=(2\pi)^3\delta^3(\p-\p')\delta_{r r'}~.
\end{align} 
	In the nonrelativistic regime relevant for the localized qubit systems considered here, the Dirac spinor reduces to
	\begin{equation}
		u_{s}(\mathbf p)
		\simeq
		\begin{pmatrix}
			\chi_{s} \\
			\dfrac{\boldsymbol{\sigma}\cdot\mathbf p}{2m_f}\chi_{s}
		\end{pmatrix}~,
	\end{equation}
	where $m_f$ is the fermion mass, \(\chi_s\) are two component Pauli spinors defined as
	\begin{equation}
		\chi_1 =
		\begin{pmatrix}
			1 \\
			0
		\end{pmatrix},
		\qquad
		\chi_2 =
		\begin{pmatrix}
			0 \\
			1
		\end{pmatrix}~.
	\end{equation}
	are the two component spin up and spin down spinors.
Also, the density operator of a the system is
	\begin{equation}\label{dno}
		\hat{\rho}=\int d\p\,\rho_{ij}(\p')\hat{c}_{i}^{\dag}(\p')\hat{c}_{j}(\p')\,,
	\end{equation}
	with the abbreviation $d\p=d^3p/(2\pi)^3$. 
	The macroscopic properties of the spin-$1/2$ particle in the interferometer are described by density matrix $\rho_{ij}$. The expectation value of the spin-$1/2$ number operator $\mathcal{\hat{D}}_{ij}(\mathbf{p})$ can be written as 
	\begin{equation}\label{efn}
		\left\langle\mathcal{\hat{D}}_{ij}(\mathbf{p})\right\rangle=\tr (\hat{\rho}\,\mathcal{\hat{D}}_{ij})=(2\pi)^{3}\delta^{3}(0)\rho_{ji}(\mathbf{p})\,.
	\end{equation}
	The density matrix $\rho_{ij}(\mathbf{p})$ of a system of spin-$1/2$ particle.
	Also, $\delta_{\sigma_0}^4(x - x(\tau))$ is a four dimensional delta function that localizes the particle at its worldline $x(\tau)$ \cite{Breuer2002}
	\begin{eqnarray}\label{eq:delta4bar}
		\delta^4_{\sigma_0}(x-\bar{x}(\tau))=\frac{1}{(2\pi \sigma^2_0)^{3/2}}\delta(x^0-\bar{x}^0(\tau))\exp\l [- \frac{(\x-\bar{\x}(\tau))^2}{2\sigma_0^2}\r]~,\end{eqnarray}
	which has a Gaussian distribution with width $\sigma_0$. 
		Using the Fourier representation of the fields and the second–order
	QED effective interaction, the interaction Hamiltonian can be written in
	momentum space as
	\begin{eqnarray}
		\hat{H}_{\textrm{int}}(x^{0}) &=&
		VV' q^{2}
		\sum_{rr'} \sum_{ss'}
		\int d\tau\, d\tau'
		\int d^{3}x\, d^{4}x'\,
		\int d\p\, d\p'\, d\q\, d\q' \;
		P_{F\,\mu\nu}(x'-x)
		\bar{u}^{A}_{r'}(\mathbf{p}')
		\,\gamma^{\mu}\,
		u^{A}_{r}(\mathbf{p})\,
		\delta_{\sigma_0}^{4}\!\big(x-\bar{x}(\tau)\big)\,
		e^{-i\,(p-p')\cdot x}
		\nonumber\\[4pt]
		&&\times\;
		\bar{u}^{B}_{s'}(\mathbf{q}')
		\,\gamma^{\nu}\,
		u^{B}_s(\mathbf{q})\,
		\delta_{\sigma_0'}^{4}\!\big(x'-\bar{x}'(\tau')\big)\,
		e^{-i\,(q-q')\cdot x'}
		c^{\dagger}_{r'}(\mathbf{p}')\,
		c_{r}(\mathbf{p})\,
		c^{\dagger}_{s'}(\mathbf{q}')\,
		c_{s}(\mathbf{q})~ .
		\label{Hmodel1}
	\end{eqnarray}
	To investigate entanglement generation, we substitute the interaction
	Hamiltonian Eq.~\eqref{Hmodel1} into the forward–scattering term of the
	generalized quantum Boltzmann equation Eq.~\eqref{QBE1} and find
	\begin{eqnarray}
		\dot{\rho}_{IJ}
		&=&
		i q^2
		\sum_{rr'}\sum_{ss'}
		\int d\tau\, d\tau'\,
		d^3x\, d^4x'\,
		d\p\, d\p'\,
		d\q\, d\q'\, dK
		\delta(x^0-\bar{x}^0(\tau))
		\delta_{\sigma_0}^{3}(\mathbf x-\bar{\mathbf x}(\tau))
		e^{-i(p^0-p'^0)x^0}
		e^{i(\mathbf p-\mathbf p')\cdot\mathbf x}
		\nonumber\\
		&&\times
		\delta_{\sigma_0'}^{4}(x'-\bar{x}'(\tau'))
		e^{-i(q^0-q'^0)x'^0}
		e^{i(\mathbf q-\mathbf q')\cdot\mathbf x'}
		\bar{u}^A_{r'}(\mathbf p')\gamma^\mu u^A_r(\mathbf p)
		\bar{u}^B_{s'}(\mathbf q')\gamma^\nu u^B_s(\mathbf q)
		P_{F\,\mu\nu}(K)
		e^{i\mathbf K\cdot(\mathbf x-\mathbf x')}
		e^{-iK^0(x^0-x'^0)}
		\nonumber\\
		&&\times
		\Big[
		\langle
		c_{r'}^\dagger(\mathbf p') c_r(\mathbf p)
		c_{s'}^\dagger(\mathbf q') c_s(\mathbf q)
		\hat{\mathcal D}_{IJ}(\mathbf k)
		\rangle
		-
		\langle
		\hat{\mathcal D}_{IJ}(\mathbf k)
		c_{r'}^\dagger(\mathbf p') c_r(\mathbf p)
		c_{s'}^\dagger(\mathbf q') c_s(\mathbf q)
		\rangle
		\Big] ,
		\label{QBE3-1}
	\end{eqnarray}
	where \(dK = d^4K/(2\pi)^4\), and we have used the standard
	normalization \(V,V'=(2\pi)^3\delta^{(3)}(0)\).
	Exploiting the independence of the spin Hilbert spaces of the two
	localized systems, the number operator can be written as
	\[
	\hat{\mathcal D}_{IJ}(\mathbf k)
	\simeq
	c_i^{A\dagger}(\mathbf k)c_j^A(\mathbf k)
	\otimes
	c_k^{B\dagger}(\mathbf k)c_l^B(\mathbf k),
	\]
	where the density matrix indices have been folded such that
	\(I\equiv (i,k)\), \(J\equiv (j,l)\) with \(i,j,k,l=1,2\).
	Substituting this expression into Eq.~\eqref{QBE3-1} gives
	\begin{eqnarray}
		\dot{\rho}_{IJ}
		&=&
		q^2
		\sum_{rr'}\sum_{ss'}
		\int d\tau\, d\tau'\,
		d^3x\, d^4x'\,
		d^3p\, d^3p'\,
		d^3q\, d^3q'\, dK
		\delta(x^0-\bar{x}^0(\tau))
		\delta_{\sigma_0}^{3}(\mathbf x-\bar{\mathbf x}(\tau))
		e^{i(\mathbf p-\mathbf p')\cdot\mathbf x}
		\nonumber\\
		&&\times
		\delta_{\sigma_0'}^{4}(x'-\bar{x}'(\tau'))
		e^{-i(q^0-q'^0)x'^0}
		e^{i(\mathbf q-\mathbf q')\cdot\mathbf x'}
		\bar{u}^A_{r'}(\mathbf p')\gamma^\mu u^A_r(\mathbf p)
		\bar{u}^B_{s'}(\mathbf q')\gamma^\nu u^B_s(\mathbf q)
		\frac{\eta_{\mu\nu}}{K^2}
		e^{-iK^0(x^0-x'^0)}
		e^{i\mathbf K\cdot(\mathbf x-\mathbf x')}
		\nonumber\\
		&&\times
		\Big[
		\langle c_{r'}^{A\dagger}(\mathbf p') c_r^A(\mathbf p)
		c_i^{A\dagger}(\mathbf k)c_j^A(\mathbf k)\rangle
		\langle c_{s'}^{B\dagger}(\mathbf q') c_s^B(\mathbf q)
		c_k^{B\dagger}(\mathbf k)c_l^B(\mathbf k)\rangle
		\nonumber\\
		&&\qquad
		-
		\langle c_i^{A\dagger}(\mathbf k)c_j^A(\mathbf k)
		c_{r'}^{A\dagger}(\mathbf p')c_r^A(\mathbf p)\rangle
		\langle c_k^{B\dagger}(\mathbf k)c_l^B(\mathbf k)
		c_{s'}^{B\dagger}(\mathbf q')c_s^B(\mathbf q)\rangle
		\Big] .
	\end{eqnarray}
	Performing the integrations over \(x\) and \(x'\) yields
	\begin{align}
		\dot{\rho}_{IJ}
		&=
		q^2
		\sum_{rr'}\sum_{ss'}
		\int d\tau\, d\tau'\,
		d^3p\, d^3p'\,
		d^3q\, d^3q'\, dK
		\frac{\eta_{\mu\nu}}{K^2}
		\bar{u}^A_{r'}(\mathbf p')\gamma^\mu u^A_r(\mathbf p)
		\bar{u}^B_{s'}(\mathbf q')\gamma^\nu u^B_s(\mathbf q)
		\nonumber\\
		&\quad\times
		\exp\!\Big[
		i(\mathbf p-\mathbf p'+\mathbf K)\cdot
		\bar{\mathbf x}(\tau)
		-i(p^0-p'^0+K^0)\bar{x}^0(\tau)
		-\frac{\sigma_0^2}{2}
		|\mathbf p-\mathbf p'+\mathbf K|^2
		\Big]
		\nonumber\\
		&\quad\times
		\exp\!\Big[
		i(\mathbf q-\mathbf q'-\mathbf K)\cdot
		\bar{\mathbf x}'(\tau')
		-i(q^0-q'^0-K^0)\bar{x}'^0(\tau')
		-\frac{\sigma_0'^2}{2}
		|\mathbf q-\mathbf q'-\mathbf K|^2
		\Big]
		\nonumber\\
		&\quad\times
		\Big[
		\langle c_{r'}^{A\dagger}(\mathbf p') c_r^A(\mathbf p)
		c_i^{A\dagger}(\mathbf k)c_j^A(\mathbf k) \rangle
		\langle c_{s'}^{B\dagger}(\mathbf q') c_s^B(\mathbf q)
		c_k^{B\dagger}(\mathbf k)c_l^B(\mathbf k) \rangle
		\nonumber\\
		&\qquad -
		\langle c_i^{A\dagger}(\mathbf k)c_j^A(\mathbf k)
		c_{r'}^{A\dagger}(\mathbf p')c_r^A(\mathbf p) \rangle
		\langle c_k^{B\dagger}(\mathbf k)c_l^B(\mathbf k)
		c_{s'}^{B\dagger}(\mathbf q')c_s^B(\mathbf q) \rangle
		\Big] .
	\end{align}
	The expectation values appearing in the above equation are evaluated
	using the standard factorized expressions
	\cite{Kosowsky:1994cy,Bavarsad:2009hm,Bartolo:2018igk,Bartolo:2019eac,
		Hoseinpour:2020hic,Sharifian:2023jem}
	\begin{align}
		\langle c^{A\dagger}_{r'}(\mathbf{p}') c^A_r(\mathbf{p})
		c_i^{A\dagger}(\mathbf{k}) c^A_j(\mathbf{k}) \rangle
		&\approx
		(2\pi)^6
		\delta^3_{\sigma_0}(\mathbf{p}-\mathbf{k})
		\delta^3_{\sigma_0}(\mathbf{p}'-\mathbf{k})
		\delta_{r i}\,
		\rho^A_{r' j}(\mathbf{k}) ,
		\\
		\langle c^{B\dagger}_{s'}(\mathbf{q}') c^B_s(\mathbf{q})
		c_k^{B\dagger}(\mathbf{k}) c^B_l(\mathbf{k}) \rangle
		&\approx
		(2\pi)^6
		\delta^3_{\sigma_0}(\mathbf{q}-\mathbf{k})
		\delta^3_{\sigma_0}(\mathbf{q}'-\mathbf{k})
		\delta_{s k}\,
		\rho^B_{s' l}(\mathbf{k}) ,
		\\
		\langle c^{A\dagger}_i(\mathbf{k}) c^A_j(\mathbf{k})
		c_{r'}^{A\dagger}(\mathbf{p}') c^A_r(\mathbf{p}) \rangle
		&\approx
		(2\pi)^6
		\delta^3_{\sigma_0}(\mathbf{p}'-\mathbf{k})
		\delta^3_{\sigma_0}(\mathbf{p}-\mathbf{k})
		\delta_{j r'}\,
		\rho^A_{i r}(\mathbf{k}) ,
		\\
		\langle c^{B\dagger}_k(\mathbf{k}) c^B_l(\mathbf{k})
		c_{s'}^{B\dagger}(\mathbf{q}') c^B_s(\mathbf{q}) \rangle
		&\approx
		(2\pi)^6
		\delta^3_{\sigma_0}(\mathbf{q}'-\mathbf{k})
		\delta^3_{\sigma_0}(\mathbf{q}-\mathbf{k})
		\delta_{l s'}\,
		\rho^B_{k s}(\mathbf{k})~ ,
	\end{align}
	where
	\begin{equation}
		\delta^3_{\sigma_0}(\mathbf p-\mathbf k)
		=
		\left(\frac{\sigma_0^2}{2\pi}\right)^{3/2}
		\exp\!\left[-\frac{\sigma_0^2}{2}
		|\mathbf p-\mathbf k|^2\right]~,
	\end{equation}
	and analogously for other delta functions.
	After performing the Gaussian integrations, the momentum
	arguments collapse to the shifted values
	\begin{equation}
		\mathbf p = \mathbf k - \frac{\mathbf K}{2}~, \qquad
		\mathbf p' = \mathbf k + \frac{\mathbf K}{2}~, \qquad
		\mathbf q = \mathbf k + \frac{\mathbf K}{2}~, \qquad
		\mathbf q' = \mathbf k - \frac{\mathbf K}{2}~.
	\end{equation}
 We begin with the spinor current contraction
	 \begin{equation}
	 	\mathcal{S}^{\mu\nu}_{rr'ss'}(\mathbf{k},\mathbf{K})
	 	=
	 	\bar{u}^A_{r'}(\mathbf{p}')
	 	\gamma^\mu
	 	u^A_r(\mathbf{p})
	 	\,
	 	\bar{u}^B_{s'}(\mathbf{q}')
	 	\gamma^\nu
	 	u^B_s(\mathbf{q})~.
	 \end{equation}
	 Contracting with the Minkowski metric and expanding the Dirac spinors
	 to order \(1/m_f^2\) gives
	 \begin{align}
	 	\eta_{\mu\nu}\mathcal{S}^{\mu\nu}_{rr'ss'}
	 	&=
	 	(\chi_{r'}^{A\dagger}\chi^A_r)
	 	(\chi_{s'}^{B\dagger}\chi^B_s)
	 	\left[
	 	1
	 	-\frac{|\mathbf k|^2}{2m_f^2}
	 	-\frac{|\mathbf K|^2}{8m_f^2}
	 	\right]
	 	\nonumber\\
	 	&\quad
	 	+
	 	\frac{|\mathbf K|^2}{4m_f^2}
	 	(\chi_{r'}^{A\dagger}\boldsymbol{\sigma}\chi^A_r)
	 	\!\cdot\!
	 	(\chi_{s'}^{B\dagger}\boldsymbol{\sigma}\chi^B_s)
	 	\nonumber\\
	 	&\quad
	 	-
	 	\frac{1}{4m_f^2}
	 	(\chi_{r'}^{A\dagger}\boldsymbol{\sigma}\chi^A_r\!\cdot\!\mathbf K)
	 	(\chi_{s'}^{B\dagger}\boldsymbol{\sigma}\chi^B_s\!\cdot\!\mathbf K)
	 	+\mathcal O(m_f^{-3})~.
	 \end{align}
	 
	 We retain only the spin-spin contribution, since the remaining terms do
	 not contribute to the entangling dynamics considered here. Introducing
	 the compact notation
	 \begin{equation}
	 	\sigma_\alpha^A(r'r)
	 	=
	 	\chi_{r'}^{A\dagger}
	 	\sigma_\alpha
	 	\chi_r^A,
	 	\qquad
	 	\sigma_\beta^B(s's)
	 	=
	 	\chi_{s'}^{B\dagger}
	 	\sigma_\beta
	 	\chi_s^B~,
	 \end{equation}
	 the spin-spin kernel becomes
	 \begin{equation}
	 	\mathcal{S}^{\mathrm{SS}}_{rr';ss'}
	 	=
	 	\frac{1}{4m_f^2}
	 	\left[
	 	\sigma_\alpha^A(r'r)\,
	 	\sigma_\beta^B(s's)\,
	 	K_\alpha K_\beta
	 	-
	 	\delta_{\alpha\beta}\,
	 	\sigma_\alpha^A(r'r)\,
	 	\sigma_\beta^B(s's)\,
	 	|\mathbf K|^2
	 	\right]~.
	 \end{equation}
	 
	 Performing the \(K^0\) integral yields
	 \begin{equation}
	 	\int \frac{dK^0}{2\pi}
	 	\frac{e^{-iK^0 t}}
	 	{(K^0)^2-|\mathbf K|^2+i0^+}
	 	=
	 	\frac{i}{2|\mathbf K|}
	 	e^{-i|\mathbf K|t}~.
	 \end{equation}
	  We now assume the long-time condition, where the observation time is much larger than the microscopic correlation time associated with the photon propagator. In this regime
	 \begin{equation}
	 	\int_{-\infty}^{+\infty} d\tau'
	 	\,e^{-i|\mathbf K|\,|t-\tau'|}
	 	=
	 	\frac{2}{|\mathbf K|}~,
	 \end{equation}
	 which generates the effective Coulomb kernel \(1/|\mathbf K|^2\).
	 The evolution equation therefore reduces to
	 \begin{align}
	 	\dot{\rho}_{IJ}
	 	&=
	 	i q^2
	 	\sum_{rr'ss'}
	 	\int\frac{d^3\mathbf K}{(2\pi)^3}
	 	\frac{e^{i\mathbf K\cdot\mathbf R}}
	 	{|\mathbf K|^2}
	 	e^{-\sigma_0^2|\mathbf K|^2}
	 	\,
	 	\mathcal{S}^{\mathrm{SS}}_{rr';ss'}
	 	\Big[
	 	\delta_{ri}\delta_{sk}\,
	 	\rho^A_{r'j}\rho^B_{s'l}
	 	-
	 	\delta_{jr'}\delta_{ls'}\,
	 	\rho^A_{ir}\rho^B_{ks}
	 	\Big]~.
	 \end{align}
	 Introducing the Coulomb kernel
	 \begin{equation}
	 	I_0(R)
	 	=
	 	\int
	 	\frac{d^3\mathbf K}{(2\pi)^3}
	 	\frac{e^{i\mathbf K\cdot\mathbf R}}
	 	{|\mathbf K|^2}
	 	e^{-\sigma_0^2|\mathbf K|^2}
	 	=
	 	\frac{1}{4\pi R}
	 	\operatorname{erf}\!\left(\frac{R}{2\sigma_0}\right)~,
	 \end{equation}
	 the evolution equation becomes
	 \begin{align}
	 	\dot{\rho}_{IJ}(\mathbf R)
	 	&=
	 	i\frac{q^2}{4m_f^2}
	 	\Big[
	 	\nabla^2 I_0(R)
	 	(\sigma^a\rho^A)_{ij}
	 	(\sigma^a\rho^B)_{kl}
	 	-
	 	\partial_a\partial_b I_0(R)
	 	(\sigma^a\rho^A)_{ij}
	 	(\sigma^b\rho^B)_{kl}
	 	\nonumber\\
	 	&\quad
	 	-
	 	\nabla^2 I_0(R)
	 	(\rho^A\sigma^a)_{ij}
	 	(\rho^B\sigma^a)_{kl}
	 	+
	 	\partial_a\partial_b I_0(R)
	 	(\rho^A\sigma^a)_{ij}
	 	(\rho^B\sigma^b)_{kl}
	 	\Big]~. \label{dotrhoIJ1}
	 \end{align}
 The derivatives of the smeared Coulomb kernel are
	 \begin{equation}
	 	\partial_i\partial_j I_0(R)
	 	=
	 	\left(\delta_{ij}-3\hat R_i\hat R_j\right)
	 	\frac{1}{R^3}
	 	\left[
	 	\operatorname{erf}\!\left(\frac{R}{2\sigma_0}\right)
	 	-
	 	\frac{2\sigma_0}{\sqrt{\pi}R}
	 	e^{-R^2/(4\sigma_0^2)}
	 	\right]~.
	 \end{equation}
	 The evolution equation contains the structure
	 \begin{equation}
	 	\sum_{rr'ss'}
	 	\mathcal S^{\mathrm{SS}}_{rr';ss'}
	 	\left[
	 	\delta_{ri}\delta_{sk}\,
	 	\rho^A_{r'j}\rho^B_{s'l}
	 	-
	 	\delta_{jr'}\delta_{ls'}\,
	 	\rho^A_{ir}\rho^B_{ks}
	 	\right]~.
	 \end{equation}
	 Using the Kronecker deltas to collapse the spin sums gives
	 \begin{align}
	 	\sum_{rr'ss'}
	 	\sigma_\alpha^A(r'r)\,
	 	\sigma_\beta^B(s's)\,
	 	\delta_{ri}\delta_{sk}\,
	 	\rho^A_{r'j}\rho^B_{s'l}
	 	&=
	 	(\sigma_\alpha\rho^A)_{ij}
	 	(\sigma_\beta\rho^B)_{kl}~,
	 	\\[2mm]
	 	\sum_{rr'ss'}
	 	\sigma_\alpha^A(r'r)\,
	 	\sigma_\beta^B(s's)\,
	 	\delta_{jr'}\delta_{ls'}\,
	 	\rho^A_{ir}\rho^B_{ks}
	 	&=
	 	(\rho^A\sigma_\alpha)_{ij}
	 	(\rho^B\sigma_\beta)_{kl}~.
	 \end{align}
		 Substituting these results into \eqref{dotrhoIJ1} yields
	 \begin{align}
	 	\dot{\rho}_{IJ}(\mathbf R)
	 	&=
	 	i\Gamma_{SS}(R)
	 	\Big[
	 	(\boldsymbol{\sigma}\rho)_{ij}
	 	\!\cdot\!
	 	(\boldsymbol{\sigma}\rho)_{kl}
	 	-
	 	3
	 	(\boldsymbol{\sigma}\rho)_{ij}
	 	\!\cdot\!
	 	\hat{\mathbf R}\,
	 	(\boldsymbol{\sigma}\rho)_{kl}
	 	\!\cdot\!
	 	\hat{\mathbf R}
	 	-
	 	(\rho\boldsymbol{\sigma})_{ij}
	 	\!\cdot\!
	 	(\rho\boldsymbol{\sigma})_{kl}
	 	+
	 	3
	 	(\rho\boldsymbol{\sigma})_{ij}
	 	\!\cdot\!
	 	\hat{\mathbf R}\,
	 	(\rho\boldsymbol{\sigma})_{kl}
	 	\!\cdot\!
	 	\hat{\mathbf R}
	 	\Big]~,
	 	\label{rhoReq2}
	 \end{align}
	 where
	 \begin{equation}
	 	\Gamma_{SS}(R)
	 	=
	 	\frac{\alpha}{4\pi m_f^2 R^3}
	 	\left[
	 	\operatorname{erf}\!\left(\frac{R}{2\sigma_0}\right)
	 	-
	 	\frac{2\sigma_0}{\sqrt{\pi}R}
	 	e^{-R^2/(4\sigma_0^2)}
	 	\right]~,
	 	\label{GammaSS1}
	 \end{equation}
	 	and in SI units, the coupling becomes
	 \begin{equation}
	 	\Gamma_{SS}(R)
	 	=
	 	\frac{\alpha\hbar^2}{4\pi m_f^2 c R^3}
	 	\left[
	 	\operatorname{erf}\!\left(\frac{R}{2\sigma_0}\right)
	 	-
	 	\frac{2\sigma_0}{\sqrt{\pi}R}
	 	e^{-R^2/(4\sigma_0^2)}
	 	\right]~.
	 \end{equation}
	 	 The evolution equation can equivalently be expressed as
	 \begin{equation}
	 	\dot{\rho}
	 	=
	 	-i
	 	\left[
	 	H_{SS},
	 	\rho
	 	\right]~,
	 \end{equation}
	 with the effective spin-spin Hamiltonian
	 \begin{equation}
	 	H_{SS}
	 	=
	 	\Gamma_{SS}(R)
	 	\left[
	 	\boldsymbol{\sigma}_1\!\cdot\!\boldsymbol{\sigma}_2
	 	-
	 	3
	 	(\boldsymbol{\sigma}_1\!\cdot\!\hat{\mathbf R})
	 	(\boldsymbol{\sigma}_2\!\cdot\!\hat{\mathbf R})
	 	\right]~.
	 \end{equation}

	\section{Three qubit interaction kernel }
	
	After inserting the Fourier representations of the fermion and photon propagators, the Hamiltonian in Eq.~\eqref{H93} becomes
	\begin{equation}
		\begin{aligned}
			\hat H_{\mathrm{int}}(t)
			&= q^4 \sum_{\substack{r_A,r'_A\\ r_{B_1},r'_{B_1}\\ r_{B_2},r'_{B_2}}}
				\int d^3x_1\, d^4x_2\, d^4x'_1\, d^4x'_2 d\tau_1\, d\tau_2\, d\tau'_1\, d\tau'_2 
			\int d\mathbf p_1 d\mathbf p_2 d\mathbf k_1 d\mathbf k'_1 d\mathbf k_2 d\mathbf k'_2
			\int dQ\, dK_1\, dK_2
			\\
			&\quad \times
			\bar u_{r'_A}(\mathbf p_1) \gamma^{\mu_1} 
			\frac{i(\slashed Q + m_A)}{Q^2 - m_A^2 + i\epsilon} 
			\gamma^{\mu_2} u_{r_A}(\mathbf p_2)
			\bar u_{r'_{B_1}}(\mathbf k'_1) \gamma^{\nu_1} u_{r_{B_1}}(\mathbf k_1)
			\;
			\bar u_{r'_{B_2}}(\mathbf k'_2) \gamma^{\nu_2} u_{r_{B_2}}(\mathbf k_2)
			\\
			&\quad \times
			\frac{-i \eta_{\mu_1 \nu_1}}{K_1^2 + i \epsilon}
			\frac{-i \eta_{\mu_2 \nu_2}}{K_2^2 + i \epsilon}
			\hat c^\dagger_{A,r'_A}(\mathbf p_1)\hat c_{A,r_A}(\mathbf p_2)
			\hat c^\dagger_{B_1,r'_{B_1}}(\mathbf k'_1)\hat c_{B_1,r_{B_1}}(\mathbf k_1)
			\hat c^\dagger_{B_2,r'_{B_2}}(\mathbf k'_2)\hat c_{B_2,r_{B_2}}(\mathbf k_2)
			\\
			&\quad \times
			e^{-i(p_1 - Q - K_1) \cdot x_1}
			e^{-i(Q - p_2 - K_2) \cdot x_2}
			e^{-i(k_1 - k'_1 +K_1) \cdot x'_1}
			e^{-i(k_2 - k'_2 + K_2) \cdot x'_2}
			\\
			&\quad \times
			\delta^{(4)}_{\sigma_A}(x_1 - \bar x_{1A}(\tau_1))
			\delta^{(4)}_{\sigma_A}(x_2 - \bar x_{2A}(\tau_2))
			\delta^{(4)}_{\sigma_{B_1}}(x'_1 - \bar x_{1B}(\tau'_1))
			\delta^{(4)}_{\sigma_{B_2}}(x'_2 - \bar x_{2B}(\tau'_2)) ~.
		\end{aligned}
	\end{equation}
	This Hamiltonian represents a fourth order effective fermionic interaction mediated by two photons, generalizing the two qubit spin-spin interaction to a three qubit configuration. After integrating out the photon degrees of freedom and projecting onto the nonrelativistic spin sector, this interaction can be mapped onto an effective spin Hamiltonian involving the qubits \((A,B_1,B_2)\).
	Using the modified quantum Boltzmann equation (QBE) for the \(8\times8\) density matrix \(\rho_{IJ}\),
	\begin{align}
		\left[(2\pi)^3\delta^3(0)\right]^3
		\dot{\rho}_{IJ}(\mathbf{k},t)
		=
		i\left\langle
		\big[\hat{H}_{\textrm{int}}(0),\hat{\mathcal{D}}_{IJ}(\mathbf{k})\big]
		\right\rangle ,
		\label{QBE-2}
	\end{align}
	where the composite indices \(I,J=1,\dots,8\) label the three qubit Hilbert space.  
	The operator \(\hat{\mathcal{D}}_{IJ}\) is constructed from the tensor product of the number density operators of subsystems \(A\), \(B_1\), and \(B_2\).
	Using the independence of the spin Hilbert spaces, the number operator can be written as
	\begin{equation}
		\hat{\mathcal{D}}_{IJ}(\mathbf k)
		\simeq
		c^{A\dagger}_i(\mathbf k)c^A_j(\mathbf k)
		\otimes
		c^{B_1\dagger}_{k_1}(\mathbf k)c^{B_1}_{l_1}(\mathbf k)
		\otimes
		c^{B_2\dagger}_{k_2}(\mathbf k)c^{B_2}_{l_2}(\mathbf k),
	\end{equation}
	where the folded density matrix indices satisfy \(i,j,k_a,l_a=1,2\).
		We now integrate over the spacetime coordinates
	\(x_1,x_2,x'_1,x'_2\) using the smeared worldline delta function \cite{Breuer2002}
	\begin{equation}
		\delta^4_{\sigma_X}(x-\bar{x}_X(\tau))
		=
		\frac{1}{(2\pi \sigma_X^2)^{3/2}}
		\delta\!\big(x^0-\bar{x}_X^0(\tau)\big)
		\exp\!\left[
		-\frac{(\mathbf{x}-\bar{\mathbf{x}}_X(\tau))^2}{2\sigma_X^2}
		\right],
	\end{equation}
	where \(X=A,B_1,B_2\) labels the corresponding subsystem.
	The temporal component is localized exactly on the worldline, while the spatial profile is Gaussian with width \(\sigma_X\). Performing the integrations over the spacetime coordinates yields
	 	\begin{align}
		\dot{\rho}_{IJ}(\mathbf{k},t)
		&=
		i q^4 V
		\sum_{\substack{r_A,r'_A\\ r_{B_1},r'_{B_1}\\ r_{B_2},r'_{B_2}}}
		\int  d\tau_1\, d\tau_2\, d\tau'_1\, d\tau'_2d\mathbf p_1\, d\mathbf p_2\, d\mathbf k_1\, d\mathbf k'_1\, d\mathbf k_2\, d\mathbf k'_2
		\int dQ\, dK_1\, dK_2 \frac{-i\eta_{\mu_1\nu_1}}{K_1^2+i\epsilon}
		\frac{-i\eta_{\mu_2\nu_2}}{K_2^2+i\epsilon}
		\nonumber\\
		&\quad\times
		\bar u_{r'_A}(\mathbf p_1)\gamma^{\mu_1}
		\frac{i(\slashed Q + m_A)}{Q^2 - m_A^2 + i\epsilon}
		\gamma^{\mu_2}u_{r_A}(\mathbf p_2)
		\bar u_{r'_{B_1}}(\mathbf k'_1)\gamma^{\nu_1}u_{r_{B_1}}(\mathbf k_1)
		\bar u_{r'_{B_2}}(\mathbf k'_2)\gamma^{\nu_2}u_{r_{B_2}}(\mathbf k_2)
		\nonumber\\
		&\quad\times
		\exp\Big[
		-i(p_1^0-Q^0-K_1^0)\bar x_{1A}^0(\tau_1)
		+i(\mathbf p_1-\mathbf Q-\mathbf K_1)\!\cdot\!\bar{\mathbf x}_{1A}(\tau_1)
		-\frac{\sigma_A^2}{2}|\mathbf p_1-\mathbf Q-\mathbf K_1|^2
		\nonumber\\
		&\qquad
		-i(Q^0-p_2^0-K_2^0)\bar x_{2A}^0(\tau_2)
		+i(\mathbf Q-\mathbf p_2-\mathbf K_2)\!\cdot\!\bar{\mathbf x}_{2A}(\tau_2)
		-\frac{\sigma_A^2}{2}|\mathbf Q-\mathbf p_2-\mathbf K_2|^2
		\nonumber\\
		&\qquad
		-i(k_1^0-k_1'^0+K_1^0)\bar x_{1B}^0(\tau'_1)
		+i(\mathbf k_1-\mathbf k'_1+\mathbf K_1)\!\cdot\!\bar{\mathbf x}_{1B}(\tau'_1)
		-\frac{\sigma_{B_1}^2}{2}|\mathbf k_1-\mathbf k'_1+\mathbf K_1|^2
		\nonumber\\
		&\qquad
		-i(k_2^0-k_2'^0+K_2^0)\bar x_{2B}^0(\tau'_2)
		+i(\mathbf k_2-\mathbf k'_2+\mathbf K_2)\!\cdot\!\bar{\mathbf x}_{2B}(\tau'_2)
		-\frac{\sigma_{B_2}^2}{2}|\mathbf k_2-\mathbf k'_2+\mathbf K_2|^2
		\Big]
			\nonumber\\
		&\quad\times
		\Big[
		\left\langle
		\hat c^\dagger_{A,r'_A}(\mathbf p_1)\hat c_{A,r_A}(\mathbf p_2)
		\hat c^\dagger_{B_1,r'_{B_1}}(\mathbf k'_1)\hat c_{B_1,r_{B_1}}(\mathbf k_1)
		\hat c^\dagger_{B_2,r'_{B_2}}(\mathbf k'_2)\hat c_{B_2,r_{B_2}}(\mathbf k_2)
		\,\hat{\mathcal D}_{IJ}(\mathbf{k})
		\right\rangle
		\nonumber\\
		&\qquad-
		\left\langle
		\hat{\mathcal D}_{IJ}(\mathbf{k})\,
		\hat c^\dagger_{A,r'_A}(\mathbf p_1)\hat c_{A,r_A}(\mathbf p_2)
		\hat c^\dagger_{B_1,r'_{B_1}}(\mathbf k'_1)\hat c_{B_1,r_{B_1}}(\mathbf k_1)
		\hat c^\dagger_{B_2,r'_{B_2}}(\mathbf k'_2)\hat c_{B_2,r_{B_2}}(\mathbf k_2)
		\right\rangle
		\Big]~.
		\label{rhodotN2-1}
	\end{align}
		where the particle energies are
	$
	p_1^0=\sqrt{|\mathbf p_1|^2+m_A^2},~
	p_2^0=\sqrt{|\mathbf p_2|^2+m_A^2},~
$$
	k_1^0=\sqrt{|\mathbf k_1|^2+m_{B_1}^2},~
	k_2^0=\sqrt{|\mathbf k_2|^2+m_{B_2}^2},~
$$
	k_1'^0=\sqrt{|\mathbf k'_1|^2+m_{B_1}^2},
	k_2'^0=\sqrt{|\mathbf k'_2|^2+m_{B_2}^2}.
$
	Using the fermionic anticommutation relations and neglecting connected
	four point correlations (Born approximation), the relevant expectation
	values reduce to products of one particle density matrices. We obtain
	\begin{align}
		\left\langle
		c^\dagger_{A,r'_A}(\mathbf p_1)c_{A,r_A}(\mathbf p_2)
		c^{A\dagger}_i(\mathbf k)c^A_j(\mathbf k)
		\right\rangle
		&\simeq
		(2\pi)^6
		\delta^{(3)}_{\sigma_A}(\mathbf p_1-\mathbf k)
		\delta^{(3)}_{\sigma_A}(\mathbf p_2-\mathbf k)
		\delta_{r_A i}\,
		\rho^A_{r'_A j}(\mathbf k)~,
		\\
		\left\langle
		c^\dagger_{B_1,r'_{B_1}}(\mathbf k'_1)c_{B_1,r_{B_1}}(\mathbf k_1)
		c^{B_1\dagger}_{i_1}(\mathbf k)c^{B_1}_{j_1}(\mathbf k)
		\right\rangle
		&\simeq
		(2\pi)^6
		\delta^{(3)}_{\sigma_{B_1}}(\mathbf k'_1-\mathbf k)
		\delta^{(3)}_{\sigma_{B_1}}(\mathbf k_1-\mathbf k)
		\delta_{r_{B_1} i_1}\,
		\rho^{B_1}_{r'_{B_1} j_1}(\mathbf k)~,
		\\
		\left\langle
		c^\dagger_{B_2,r'_{B_2}}(\mathbf k'_2)c_{B_2,r_{B_2}}(\mathbf k_2)
		c^{B_2\dagger}_{i_2}(\mathbf k)c^{B_2}_{j_2}(\mathbf k)
		\right\rangle
		&\simeq
		(2\pi)^6
		\delta^{(3)}_{\sigma_{B_2}}(\mathbf k'_2-\mathbf k)
		\delta^{(3)}_{\sigma_{B_2}}(\mathbf k_2-\mathbf k)
		\delta_{r_{B_2} i_2}\,
		\rho^{B_2}_{r'_{B_2} j_2}(\mathbf k)~,
		\\
		\left\langle
		c^{A\dagger}_i(\mathbf k)c^A_j(\mathbf k)
		c^\dagger_{A,r'_A}(\mathbf p_1)c_{A,r_A}(\mathbf p_2)
		\right\rangle
		&\simeq
		(2\pi)^6
		\delta^{(3)}_{\sigma_A}(\mathbf p_1-\mathbf k)
		\delta^{(3)}_{\sigma_A}(\mathbf p_2-\mathbf k)
		\delta_{j r'_A}\,
		\rho^A_{i r_A}(\mathbf k)~,
		\\
		\left\langle
		c^{B_1\dagger}_{i_1}(\mathbf k)c^{B_1}_{j_1}(\mathbf k)
		c^\dagger_{B_1,r'_{B_1}}(\mathbf k'_1)c_{B_1,r_{B_1}}(\mathbf k_1)
		\right\rangle
		&\simeq
		(2\pi)^6
		\delta^{(3)}_{\sigma_{B_1}}(\mathbf k'_1-\mathbf k)
		\delta^{(3)}_{\sigma_{B_1}}(\mathbf k_1-\mathbf k)
		\delta_{j_1 r'_{B_1}}\,
		\rho^{B_1}_{i_1 r_{B_1}}(\mathbf k)~,
		\\
		\left\langle
		c^{B_2\dagger}_{i_2}(\mathbf k)c^{B_2}_{j_2}(\mathbf k)
		c^\dagger_{B_2,r'_{B_2}}(\mathbf k'_2)c_{B_2,r_{B_2}}(\mathbf k_2)
		\right\rangle
		&\simeq
		(2\pi)^6
		\delta^{(3)}_{\sigma_{B_2}}(\mathbf k'_2-\mathbf k)
		\delta^{(3)}_{\sigma_{B_2}}(\mathbf k_2-\mathbf k)
		\delta_{j_2 r'_{B_2}}\,
		\rho^{B_2}_{i_2 r_{B_2}}(\mathbf k)~.
	\end{align}
		We assume an initially factorized density matrix and neglect
	higher order connected correlations (Born approximation). The total
	density matrix therefore factorizes into independent spin Hilbert
	spaces associated with subsystems \(A\), \(B_1\), and \(B_2\),
	\begin{equation}
		\rho_{\mathrm{tot}}
		\simeq
		\rho^A \otimes \rho^{B_1} \otimes \rho^{B_2}~.
	\end{equation}
	We further assume that the momentum distributions of each subsystem are
	sharply peaked around a common momentum \(\mathbf{k}\), with finite
	spatial resolution characterized by Gaussian widths
	\(\sigma_A\), \(\sigma_{B_1}\), and \(\sigma_{B_2}\).
	The Gaussian smearing of the particle worldlines induces smeared
	momentum delta functions. For each species \(X = A, B_1, B_2\), we define
	\begin{equation}
		\delta^{(3)}_{\sigma_X}(\mathbf p-\mathbf k)
		=
		\left(\frac{\sigma_X^2}{2\pi}\right)^{3/2}
		\exp\!\left[
		-\frac{\sigma_X^2}{2}(\mathbf p-\mathbf k)^2
		\right],
		\qquad X=A,B_1,B_2~ .
	\end{equation}
After performing the momentum integrations and working in the nonrelativistic limit, the evolution equation becomes
   \begin{align}
   	\dot{\rho}_{IJ}(\mathbf{k},t)
   	&= i q^4 V
   	\sum_{\substack{r_A,r'_A\\ r_{B_1},r'_{B_1}\\ r_{B_2},r'_{B_2}}}
   	\int \frac{d^4Q}{(2\pi)^4}
   	\frac{d^4K_1}{(2\pi)^4}
   	\frac{d^4K_2}{(2\pi)^4}
   	\int d\tau_1\, d\tau_2\, d\tau'_1\, d\tau'_2 \delta\!\big(x_{1A}^0-\bar x_{1A}^0(\tau_1)\big)
   	\nonumber\\
   	&\quad\times
   		\frac{-i \eta_{\mu_1\nu_1}}{K_1^2+i\epsilon}
   	\frac{-i \eta_{\mu_2\nu_2}}{K_2^2+i\epsilon}
   	\bar u_{r'_A}\!\left(\frac{\mathbf k + \mathbf Q + \mathbf K_1}{2}\right)
   	\gamma^{\mu_1}
   	\frac{i(\slashed Q + m_A)}{Q^2 - m_A^2 + i\epsilon}
   	\gamma^{\mu_2}
   	u_{r_A}\!\left(\frac{\mathbf k + \mathbf Q - \mathbf K_2}{2}\right)
   	\nonumber\\
   	&\quad\times
   	\bar u_{r'_{B_1}}\!\left(\mathbf k+\frac{\mathbf K_1}{2}\right)
   	\gamma^{\nu_1}
   	u_{r_{B_1}}\!\left(\mathbf k-\frac{\mathbf K_1}{2}\right)
   	\bar u_{r'_{B_2}}\!\left(\mathbf k+\frac{\mathbf K_2}{2}\right)
   	\gamma^{\nu_2}
   	u_{r_{B_2}}\!\left(\mathbf k-\frac{\mathbf K_2}{2}\right)
   	\nonumber\\
   	&\quad\times
   	\exp\!\Big[
   	-i(m_A-Q^0-K_1^0)x_{1A}^0
   	-i(Q^0-m_A-K_2^0)\bar x_{2A}^0(\tau_2)
   	\Big]
   	\nonumber\\
   	&\quad\times
   	\exp\!\Big[
   	-iK_1^0\bar x_{1B}^0(\tau'_1)
   	-iK_2^0\bar x_{2B}^0(\tau'_2)
   	\Big]
   	\nonumber\\
   	&\quad\times
   	\exp\!\Big[
   	-i\mathbf Q\cdot(\bar{\mathbf x}_{1A}-\bar{\mathbf x}_{2A})
   	-i\mathbf K_1\cdot(\bar{\mathbf x}_{1A}-\bar{\mathbf x}_{1B})
   	-i\mathbf K_2\cdot(\bar{\mathbf x}_{2A}-\bar{\mathbf x}_{2B})
   	\Big]
   	\nonumber\\
   	&\quad\times
   	\exp\!\left[
   	-\frac{\sigma_A^2}{4}(|\mathbf K_1|^2+|\mathbf K_2|^2)
   	-\frac{\sigma_{B_1}^2}{2}|\mathbf K_1|^2
   	-\frac{\sigma_{B_2}^2}{2}|\mathbf K_2|^2
   	\right]
   	\nonumber\\
   	&\quad\times
   	\Big[
   	\delta_{r_A i}\delta_{r_{B_1} i_1}\delta_{r_{B_2} i_2}
   	\rho^A_{r'_A j}(\mathbf k)
   	\rho^{B_1}_{r'_{B_1} j_1}(\mathbf k)
   	\rho^{B_2}_{r'_{B_2} j_2}(\mathbf k)
   	\nonumber\\
   	&\qquad -
   	\delta_{j r'_A}\delta_{j_1 r'_{B_1}}\delta_{j_2 r'_{B_2}}
   	\rho^A_{i r_A}(\mathbf k)
   	\rho^{B_1}_{i_1 r_{B_1}}(\mathbf k)
   	\rho^{B_2}_{i_2 r_{B_2}}(\mathbf k)
   	\Big]~.
   	\label{rhodotN2-2}
   \end{align}
We now perform the contour integration over $Q^0$ and using the nonrelativistic spinor
approximation
\[
\bar u_{r'_A}\!\left(\tfrac{\mathbf k + \mathbf Q + \mathbf K_1}{2}\right)
\approx \bar u_{r'_A}(\mathbf k)~, 
\qquad
u_{r_A}\!\left(\tfrac{\mathbf k + \mathbf Q - \mathbf K_2}{2}\right)
\approx u_{r_A}(\mathbf k)~,
\]
valid for $|\mathbf Q|,|\mathbf K_i|\ll |\mathbf k|,m_A$, we obtain
\begin{align}
	\dot{\rho}_{IJ}(\mathbf{k},t)
	&=
	i q^4 V 
	\sum_{\substack{r_A,r'_A\\ r_{B_1},r'_{B_1}\\ r_{B_2},r'_{B_2}}}
	\int \frac{d^3\mathbf Q}{(2\pi)^3}
	\int \frac{d^4K_1}{(2\pi)^4}
	\int \frac{d^4K_2}{(2\pi)^4}
	\int d\tau_1\, d\tau_2\, d\tau'_1\, d\tau'_2\;
	\delta\!\big(x_{1A}^0-\bar x_{1A}^0(\tau_1)\big)
	\nonumber\\
	&\quad\times
	\bar u_{r'_A}(\mathbf k)
	\gamma^{\mu_1}
	\frac{1}{2\omega_{\mathbf Q}}
	\Big[
	(\gamma^0\omega_{\mathbf Q}
	-\boldsymbol{\gamma}\!\cdot\!\mathbf Q+m_A)
	\theta(\Delta t_A)e^{-i\omega_{\mathbf Q}\Delta t_A}
	+
	(-\gamma^0\omega_{\mathbf Q}
	-\boldsymbol{\gamma}\!\cdot\!\mathbf Q+m_A)
	\theta(-\Delta t_A)e^{+i\omega_{\mathbf Q}\Delta t_A}
	\Big]
	\gamma^{\mu_2}
	u_{r_A}(\mathbf k)
	\nonumber\\
	&\quad\times
	\frac{-i\eta_{\mu_1\nu_1}}{K_1^2+i\epsilon}
	\frac{-i\eta_{\mu_2\nu_2}}{K_2^2+i\epsilon}
	\bar u_{r'_{B_1}}\!\left(\mathbf k+\tfrac{\mathbf K_1}{2}\right)
	\gamma^{\nu_1}
	u_{r_{B_1}}\!\left(\mathbf k-\tfrac{\mathbf K_1}{2}\right)
	\;
	\bar u_{r'_{B_2}}\!\left(\mathbf k+\tfrac{\mathbf K_2}{2}\right)
	\gamma^{\nu_2}
	u_{r_{B_2}}\!\left(\mathbf k-\tfrac{\mathbf K_2}{2}\right)
	\nonumber\\
	&\quad\times
	\exp\!\Big[
	-i(m_A-K_1^0)x_{1A}^0
	-i(m_A-K_2^0)\bar x_{2A}^0(\tau_2)
	-iK_1^0\bar x_{1B}^0(\tau'_1)
	-iK_2^0\bar x_{2B}^0(\tau'_2)
	\notag\\
	&\qquad
	-i\mathbf Q\!\cdot\!(\bar{\mathbf x}_{1A}-\bar{\mathbf x}_{2A})
	-i\mathbf K_1\!\cdot\!(\bar{\mathbf x}_{1A}-\bar{\mathbf x}_{1B})
	-i\mathbf K_2\!\cdot\!(\bar{\mathbf x}_{2A}-\bar{\mathbf x}_{2B})
	\Big]
	\nonumber\\
	&\quad\times
	\exp\!\left[
	-\frac{\sigma_A^2}{4}\!\left(|\mathbf K_1|^2+|\mathbf K_2|^2\right)
	-\frac{\sigma_{B_1}^2}{2}|\mathbf K_1|^2
	-\frac{\sigma_{B_2}^2}{2}|\mathbf K_2|^2
	\right]
	\nonumber\\
	&\quad\times
	\Big[
	\delta_{r_A i}\,\delta_{r_{B_1} i_1}\,\delta_{r_{B_2} i_2}\;
	\rho^A_{r'_A j}(\mathbf k)\,
	\rho^{B_1}_{r'_{B_1} j_1}(\mathbf k)\,
	\rho^{B_2}_{r'_{B_2} j_2}(\mathbf k)
	-
	\delta_{j r'_A}\,\delta_{j_1 r'_{B_1}}\,\delta_{j_2 r'_{B_2}}\;
	\rho^A_{i r_A}(\mathbf k)\,
	\rho^{B_1}_{i_1 r_{B_1}}(\mathbf k)\,
	\rho^{B_2}_{i_2 r_{B_2}}(\mathbf k)
	\Big]~.
	\label{rhodotN2-NR}
\end{align}
We assume that the fermion $A$ is nonrelativistic,
\begin{equation}
	\omega_{\mathbf Q} \simeq m_A + \frac{\mathbf Q^2}{2m_A}~,
	\qquad
	\frac{\gamma^0 \omega_{\mathbf Q} - \boldsymbol{\gamma}\!\cdot\!\mathbf Q + m_A}{2\omega_{\mathbf Q}}
	\simeq \Lambda_+ \equiv \frac{1+\gamma^0}{2}~,
\end{equation}
which projects onto the positive–energy particle subspace.
The remaining Gaussian integral over $\mathbf Q$ gives the
nonrelativistic propagator of particle $A$,
\begin{equation}
	G_A(\Delta \mathbf x_A,\Delta t_A)
	=
	\int \frac{d^3\mathbf Q}{(2\pi)^3}
	\exp\!\left[
	-i\frac{\mathbf Q^2}{2m_A}\Delta t_A
	-i\mathbf Q\!\cdot\!\Delta\mathbf x_A
	\right]
	=
	\left(\frac{m_A}{2\pi i \Delta t_A}\right)^{3/2}
	\exp\!\left[
	i\frac{m_A |\Delta\mathbf x_A|^2}{2\Delta t_A}
	\right]~,
\end{equation}
where $\Delta\mathbf x_A = \bar{\mathbf x}_{1A}-\bar{\mathbf x}_{2A}$.
Substituting this result, the evolution equation becomes
\begin{align}
	\dot{\rho}_{IJ}(\mathbf k,t)
	&=
	i q^4 V
	\sum_{\substack{r_A,r'_A\\ r_{B_1},r'_{B_1}\\ r_{B_2},r'_{B_2}}}
	\int \frac{d^4K_1}{(2\pi)^4}
	\int \frac{d^4K_2}{(2\pi)^4}
	\int d\tau_1\, d\tau_2\, d\tau'_1\, d\tau'_2\;
	\delta\!\big(x_{1A}^0 - \bar x_{1A}^0(\tau_1)\big)
	\frac{-i\,\eta_{\mu_1\nu_1}}{K_1^2 + i\epsilon}
	\frac{-i\,\eta_{\mu_2\nu_2}}{K_2^2 + i\epsilon}
	G_A(\Delta\mathbf x_A,\Delta t_A)
	\nonumber\\
	&\quad\times
	\bar u_{r'_A}(\mathbf k)\,
	\gamma^{\mu_1}\Lambda_+\gamma^{\mu_2}\,
	u_{r_A}(\mathbf k)
	\;
	\bar u_{r'_{B_1}}\!\left(\mathbf k+\tfrac{\mathbf K_1}{2}\right)
	\gamma^{\nu_1}
	u_{r_{B_1}}\!\left(\mathbf k-\tfrac{\mathbf K_1}{2}\right)
	\;
	\bar u_{r'_{B_2}}\!\left(\mathbf k+\tfrac{\mathbf K_2}{2}\right)
	\gamma^{\nu_2}
	u_{r_{B_2}}\!\left(\mathbf k-\tfrac{\mathbf K_2}{2}\right)
	\nonumber\\
	&\quad\times
	\exp\!\Big[
	-i (m_A - K_1^0) x_{1A}^0
	-i (m_A - K_2^0)\bar x_{2A}^0(\tau_2)
	-i K_1^0 \bar x_{1B}^0(\tau'_1)
	-i K_2^0 \bar x_{2B}^0(\tau'_2)
	\nonumber\\
	&\qquad\quad
	-i\mathbf K_1\!\cdot\!(\bar{\mathbf x}_{1A}-\bar{\mathbf x}_{1B})
	-i\mathbf K_2\!\cdot\!(\bar{\mathbf x}_{2A}-\bar{\mathbf x}_{2B})
	\Big]
	\nonumber\\
	&\quad\times
	\exp\!\left[
	-\frac{\sigma_A^2}{4}(|\mathbf K_1|^2+|\mathbf K_2|^2)
	-\frac{\sigma_{B_1}^2}{2}|\mathbf K_1|^2
	-\frac{\sigma_{B_2}^2}{2}|\mathbf K_2|^2
	\right]
	\nonumber\\
	&\quad\times
	\Big[
	\delta_{r_A i}\delta_{r_{B_1} i_1}\delta_{r_{B_2} i_2}\;
	\rho^A_{r'_A j}(\mathbf k)\,
	\rho^{B_1}_{r'_{B_1} j_1}(\mathbf k)\,
	\rho^{B_2}_{r'_{B_2} j_2}(\mathbf k)
	-
	\delta_{j r'_A}\delta_{j_1 r'_{B_1}}\delta_{j_2 r'_{B_2}}\;
	\rho^A_{i r_A}(\mathbf k)\,
	\rho^{B_1}_{i_1 r_{B_1}}(\mathbf k)\,
	\rho^{B_2}_{i_2 r_{B_2}}(\mathbf k)
	\Big]~,
\end{align}
where $\Delta t_A \equiv x_{1A}^0 - \bar x_{2A}^0(\tau_2)$.
Now, using the nonrelativistic identification of time variables
\begin{equation}
	\bar{x}^0_{1A}(\tau_1)=\tau_1 , \qquad \bar{x}'^{0}_{1B}(\tau'_1)=\tau'_1~,
	\qquad \bar{x}'^{0}_{2A}(\tau_2)=\tau_2~,
	\qquad \bar{x}'^{0}_{2B}(\tau'_2)=\tau'_2~.
\end{equation}
Tthe density matrix evolution after $K^0$ integration becomes
\begin{align}
	\dot{\rho}_{IJ}(\mathbf k,t)
	&=
	i q^4 V
	\sum_{\substack{r_A,r'_A\\ r_{B_1},r'_{B_1}\\ r_{B_2},r'_{B_2}}}
	\int \frac{d^3K_1}{(2\pi)^3}
	\int \frac{d^4K_2}{(2\pi)^4}
	\int d\tau_1\, d\tau_2\, d\tau'_1\, d\tau'_2\;
	\delta(t - \tau_1)
	\frac{1}{2\omega_1}
	\frac{-i\,\eta_{\mu_2\nu_2}}{K_2^2+i\epsilon} I_{K^0_1}
	G_A(\Delta\mathbf x_A,\Delta t_A)
	\nonumber\\
	&\quad\times
	\bar u_{r'_A}(\mathbf k)\,
	\gamma^{\mu_1}\Lambda_+\gamma^{\mu_2}\,
	u_{r_A}(\mathbf k)\,
	\bar u_{r'_{B_1}}\!\left(\mathbf k+\tfrac{\mathbf K_1}{2}\right)
	\gamma^{\nu_1}
	u_{r_{B_1}}\!\left(\mathbf k-\tfrac{\mathbf K_1}{2}\right)
	\;
	\bar u_{r'_{B_2}}\!\left(\mathbf k+\tfrac{\mathbf K_2}{2}\right)
	\gamma^{\nu_2}
	u_{r_{B_2}}\!\left(\mathbf k-\tfrac{\mathbf K_2}{2}\right)
	\nonumber\\
	&\quad\times
	\exp\!\Big[
	-im_A \tau_1
	+i(m_A+K_2^0)\tau_2
	-iK_2^0\tau'_2
	-i\mathbf K_1\!\cdot\!(\bar{\mathbf x}_{1A}-\bar{\mathbf x}_{1B})
	-i\mathbf K_2\!\cdot\!(\bar{\mathbf x}_{2A}-\bar{\mathbf x}_{2B})
	\Big]
	\nonumber\\
	&\quad\times
	\exp\!\left[
	-\frac{\sigma_A^2}{4}(|\mathbf K_1|^2+|\mathbf K_2|^2)
	-\frac{\sigma_{B_1}^2}{2}|\mathbf K_1|^2
	-\frac{\sigma_{B_2}^2}{2}|\mathbf K_2|^2
	\right]
	\nonumber\\
	&\quad\times
	\Big[
	\delta_{r_A i}\delta_{r_{B_1} i_1}\delta_{r_{B_2} i_2}\;
	\rho^A_{r'_A j}(\mathbf k)\,
	\rho^{B_1}_{r'_{B_1} j_1}(\mathbf k)\,
	\rho^{B_2}_{r'_{B_2} j_2}(\mathbf k)
	-
	\delta_{j r'_A}\delta_{j_1 r'_{B_1}}\delta_{j_2 r'_{B_2}}\;
	\rho^A_{i r_A}(\mathbf k)\,
	\rho^{B_1}_{i_1 r_{B_1}}(\mathbf k)\,
	\rho^{B_2}_{i_2 r_{B_2}}(\mathbf k)
	\Big]~,
\end{align}
where $\omega_1 \equiv |\mathbf K_1|$, $\Delta t_A \equiv x_{1A}^0 - \bar x_{2A}^0(\tau_2)$.
Also, the integrate over $K_2^0$ gives
\begin{align}
	I_{K_2^0}
	&=
	\frac{1}{2\omega_2}
	\left[
	\theta(\Delta t_2)e^{-i\omega_2\Delta t_2}
	+
	\theta(-\Delta t_2)e^{+i\omega_2\Delta t_2}
	\right]~,
\end{align}
where $\omega_2 \equiv |\mathbf K_2| $ and $\Delta t_2 \equiv \tau_2-\tau'_2$.
Finally, we find
\begin{align}
	\dot{\rho}_{IJ}(\mathbf{k},t)
	&=
	i q^4 V
	\sum_{\substack{r_A,r'_A\\ r_{B_1},r'_{B_1}\\ r_{B_2},r'_{B_2}}}
	\int \frac{d^3\mathbf K_1}{(2\pi)^3}
	\int \frac{d^3\mathbf K_2}{(2\pi)^3}
	\int d\tau_1\, d\tau_2\, d\tau'_1\, d\tau'_2
	\;
	\delta(t-\tau_1)
	\frac{\eta_{\mu_1\nu_1}}{2\omega_1}
	\frac{\eta_{\mu_2\nu_2}}{2\omega_2}
	I_{K_1^0} I_{K_2^0}
	G_A(\Delta\mathbf x_A,\Delta t_A)
	\nonumber\\
	&\quad\times
	\bar u_{r'_A}(\mathbf k)
	\gamma^{\mu_1}\Lambda_+\gamma^{\mu_2}
	u_{r_A}(\mathbf k)
	\bar u_{r'_{B_1}}\!\left(\mathbf k+\frac{\mathbf K_1}{2}\right)
	\gamma^{\nu_1}
	u_{r_{B_1}}\!\left(\mathbf k-\frac{\mathbf K_1}{2}\right)
	\bar u_{r'_{B_2}}\!\left(\mathbf k+\frac{\mathbf K_2}{2}\right)
	\gamma^{\nu_2}
	u_{r_{B_2}}\!\left(\mathbf k-\frac{\mathbf K_2}{2}\right)
	\nonumber\\
	&\quad\times
	\exp\!\Big[
	-im_A t
	+im_A \tau_2
	-i\mathbf K_1\!\cdot\!(\bar{\mathbf x}_{1A}-\bar{\mathbf x}_{1B})
	-i\mathbf K_2\!\cdot\!(\bar{\mathbf x}_{2A}-\bar{\mathbf x}_{2B})
	\Big]
	\nonumber\\
	&\quad\times
	\exp\!\left[
	-\frac{\sigma_A^2}{4}(|\mathbf K_1|^2+|\mathbf K_2|^2)
	-\frac{\sigma_{B_1}^2}{2}|\mathbf K_1|^2
	-\frac{\sigma_{B_2}^2}{2}|\mathbf K_2|^2
	\right]
	\nonumber\\
	&\quad\times
	\Big[
	\delta_{r_A i}\delta_{r_{B_1}k_1}\delta_{r_{B_2}k_2}
	\rho^A_{r'_A j}(\mathbf k)
	\rho^{B_1}_{r'_{B_1}l_1}(\mathbf k)
	\rho^{B_2}_{r'_{B_2}l_2}(\mathbf k)
-
	\delta_{j r'_A}\delta_{l_1 r'_{B_1}}\delta_{l_2 r'_{B_2}}
	\rho^A_{i r_A}(\mathbf k)
	\rho^{B_1}_{k_1 r_{B_1}}(\mathbf k)
	\rho^{B_2}_{k_2 r_{B_2}}(\mathbf k)
	\Big]~.
\end{align}
Performing the $\tau_1$ integral sets $\tau_1 = t$.  
The remaining proper time integrations are dominated in the 
Markovian limit, $T\gg1/\omega_{1,2}$, by the slowly varying
envelope.  Using
\begin{equation}
\int d\tau'_1\, I_{K_1^0} \simeq \frac{1}{\omega_1^2}~, 
\qquad
\int d\tau'_2\, I_{K_2^0} \simeq \frac{1}{\omega_2^2}~,
\end{equation}
and the closure relation of the nonrelativistic propagator
\begin{equation}
\int_0^{\infty} d\tau_2\;
G_A(\Delta\mathbf x_A,t-\tau_2)\,e^{-im_A(t-\tau_2)}
=
\frac{i}{m_A}\,\delta^3(\Delta\mathbf x_A)~,
\end{equation}
we obtain
\begin{align}
	\dot{\rho}_{IJ}(\mathbf k,t)
	&=
	-i\,q^4\,V\,\frac{\delta^3(\Delta\mathbf x_A)}{m_A}
	\sum_{\substack{r_A,r'_A\\ r_{B_1},r'_{B_1}\\ r_{B_2},r'_{B_2}}}
	\int\frac{d^3\mathbf K_1}{(2\pi)^3}
	\int\frac{d^3\mathbf K_2}{(2\pi)^3}
	\frac{\eta_{\mu_1\nu_1}}{2\omega_1^2}
	\frac{\eta_{\mu_2\nu_2}}{2\omega_2^2}	\bar u_{r'_A}(\mathbf k)\,
	\gamma^{\mu_1}\Lambda_+\gamma^{\mu_2}\,
	u_{r_A}(\mathbf k)
	\nonumber\\
	&\quad\times
	\;\bar u_{r'_{B_1}}\!\left(\mathbf k+\frac{\mathbf K_1}{2}\right)
	\gamma^{\nu_1}
	u_{r_{B_1}}\!\left(\mathbf k-\frac{\mathbf K_1}{2}\right)
	\bar u_{r'_{B_2}}\!\left(\mathbf k+\frac{\mathbf K_2}{2}\right)
	\gamma^{\nu_2}
	u_{r_{B_2}}\!\left(\mathbf k-\frac{\mathbf K_2}{2}\right)
	\nonumber\\
	&\quad\times
	\exp\!\Big[
	-i\mathbf K_1\!\cdot(\bar{\mathbf x}_{1A}-\bar{\mathbf x}_{1B})
	-i\mathbf K_2\!\cdot(\bar{\mathbf x}_{2A}-\bar{\mathbf x}_{2B})
	\Big]
	\exp\!\left[
	-\tfrac{\sigma_A^2}{4}(|\mathbf K_1|^2+|\mathbf K_2|^2)
	-\tfrac{\sigma_{B_1}^2}{2}|\mathbf K_1|^2
	-\tfrac{\sigma_{B_2}^2}{2}|\mathbf K_2|^2
	\right]
	\nonumber\\
	&\quad\times
	\Big[
	\delta_{r_A i}\delta_{r_{B_1}k_1}\delta_{r_{B_2}k_2}
	\rho^A_{r'_A j}(\mathbf k)
	\rho^{B_1}_{r'_{B_1}l_1}(\mathbf k)
	\rho^{B_2}_{r'_{B_2}l_2}(\mathbf k)
	\;-\;
	\delta_{j r'_A}\delta_{l_1 r'_{B_1}}\delta_{l_2 r'_{B_2}}
	\rho^A_{i r_A}(\mathbf k)
	\rho^{B_1}_{k_1 r_{B_1}}(\mathbf k)
	\rho^{B_2}_{k_2 r_{B_2}}(\mathbf k)
	\Big]~.
\end{align}

\subsection{Nonrelativistic expansion of the three fermion amplitude}

We derive here the nonrelativistic expansion of the three fermion exchange
amplitude, retaining all terms through $\mathcal{O}(1/m^2)$ that contribute to
scalar, spin-orbit, and spin-spin interactions.
The amplitude takes the form
\begin{equation}
	\mathcal{M}
	=
	\frac{\eta_{\mu_1\nu_1}}{2\omega_1^2}
	\frac{\eta_{\mu_2\nu_2}}{2\omega_2^2}
	\,J_A^{\mu_1\mu_2}\,
	J_{B_1}^{\nu_1}\,
	J_{B_2}^{\nu_2}~,
\end{equation}
where
\begin{equation}
	J_A^{\mu_1\mu_2}
	=
	\bar u_{A'}(\mathbf k)\,
	\gamma^{\mu_1}\Lambda_A\gamma^{\mu_2}\,
	u_A(\mathbf k)~,
	\qquad
	\Lambda_A = \frac{1+\gamma^0}{2}~,
\end{equation}
and
\begin{equation}
	J_{B_i}^{\nu}
	=
	\bar u_{B_i'}\!\left(\mathbf k+\frac{\mathbf K_i}{2}\right)
	\gamma^\nu\,
	u_{B_i}\!\left(\mathbf k-\frac{\mathbf K_i}{2}\right)~,
	\qquad i=1,2~.
\end{equation}
We work throughout in the nonrelativistic domain
\begin{equation}
	|\mathbf k|,\,|\mathbf K_i| \ll m_A,\,m_{B_i}~.
\end{equation}
Standard Pauli--Dirac reduction then yields the following.
Since the projector $\Lambda_A$ eliminates lower components,
$A$ contributes only through positive energy Pauli structures.
To leading nonrelativistic order
\begin{align}
	J_A^{00}
	&= \chi_A'^{\!\dagger}\chi_A~,
	\\
	J_A^{0i} = J_A^{i0}
	&=
	\frac{k^i}{m_A}
	\left(\chi_A'^{\!\dagger}\chi_A\right)
	+
	\frac{i}{2m_A}\,
	\left(\mathbf k\times
	\chi_A'^{\!\dagger}\boldsymbol{\sigma}\chi_A\right)^i~,
	\\
	J_A^{ij}
	&=
	\delta^{ij}\,
	\left(\chi_A'^{\!\dagger}\chi_A\right)
	+
	\mathcal{O}\!\left(\frac{|\mathbf k|}{m_A}\right)~.
\end{align}
Expanding to second order in momenta gives the time component
\begin{align}
	J_{B_i}^0
	&=
	\left(\chi_{B_i'}^{\!\dagger}\chi_{B_i}\right)
	\left[
	1
	-\frac{k^2}{2m_{B_i}^2}
	-\frac{K_i^2}{8m_{B_i}^2}
	\right]
	+
	\frac{i}{4m_{B_i}^2}\,
	\chi_{B_i'}^{\!\dagger}
	\left( (\mathbf k\times\mathbf K_i)\cdot\boldsymbol{\sigma}\right)
	\chi_{B_i}~,
\end{align}
and the spatial component
\begin{align}
	J_{B_i}^j
	&=
	\frac{k^j}{m_{B_i}}\,
	\left(\chi_{B_i'}^{\!\dagger}\chi_{B_i}\right)
	-
	\frac{i}{2m_{B_i}}\,
	\chi_{B_i'}^{\!\dagger}
	\left(\mathbf K_i\times\boldsymbol{\sigma}\right)^j
	\chi_{B_i}~.
\end{align}
These expressions supply all scalar, spin-orbit,
and spin-spin structures that enter the three body kernel.
The scalar term, $\mathcal{S}_0$, arising from $J_A^{00}J_{B_1}^0J_{B_2}^0$ is
\begin{align}
	\mathcal{S}_0
	&=
	(\chi_A'^\dagger \chi_A)
	(\chi_{B_1'}^\dagger \chi_{B_1})
	(\chi_{B_2'}^\dagger \chi_{B_2})
	\left[
	1
	- \frac{|\k|^2}{2 m_{B_1}^2}
	- \frac{|\k|^2}{2 m_{B_2}^2}
	- \frac{|\K_1|^2}{8 m_{B_1}^2}
	- \frac{|\K_2|^2}{8 m_{B_2}^2}
	\right]~.
\end{align}
The spin-orbit terms, $\mathcal{S}_{\mathrm{SO}}^{(1)}$ and $\mathcal{S}_{\mathrm{SO}}^{(2)}$, arise from the cross terms $J_A^{0i} J_{B_1}^i J_{B_2}^0$ and $J_A^{i0} J_{B_1}^0 J_{B_2}^i$ take the forms
\begin{align}
	\mathcal{S}_{\mathrm{SO}}^{(1)}
	&=
	(\chi_A'^\dagger \chi_A)
	(\chi_{B_2'}^\dagger \chi_{B_2})
	\, i
	(\chi_{B_1'}^\dagger \boldsymbol{\sigma} \chi_{B_1})
	\cdot
	(\mathbf{k} \times \mathbf{K}_1)
	\left[
	\frac{1}{4 m_{B_1}^2}
	-
	\frac{1}{m_A m_{B_1}}
	\right]~,
	\\
	\mathcal{S}_{\mathrm{SO}}^{(2)}
	&=
	(\chi_A'^\dagger \chi_A)
	(\chi_{B_1'}^\dagger \chi_{B_1})
	\, i
	(\chi_{B_2'}^\dagger \boldsymbol{\sigma} \chi_{B_2})
	\cdot
	(\mathbf{k} \times \mathbf{K}_2)
	\left[
	\frac{1}{4 m_{B_2}^2}
	-
	\frac{1}{m_A m_{B_2}}
	\right]~.
\end{align}
The spin-spin interaction arises from the spatial contraction
$J_A^{ij} J_{B_1}^i J_{B_2}^j$. Using the leading contribution
$J_A^{ij} = (\chi_A'^\dagger \chi_A)\delta^{ij}$, we obtain
\begin{align}
	\mathcal{S}_{\mathrm{SS}}
	&=
	-
	\frac{1}{4 m_{B_1} m_{B_2}}
	(\chi_A'^\dagger \chi_A)
	\Big[
	(\chi_{B_1'}^\dagger \boldsymbol{\sigma} \chi_{B_1})
	\cdot
	(\chi_{B_2'}^\dagger \boldsymbol{\sigma} \chi_{B_2})
	\, (\mathbf{K}_1 \cdot \mathbf{K}_2)
	-
	(\chi_{B_1'}^\dagger \boldsymbol{\sigma} \cdot \mathbf{K}_2 \, \chi_{B_1})
	(\chi_{B_2'}^\dagger \boldsymbol{\sigma} \cdot \mathbf{K}_1 \, \chi_{B_2})
	\Big]~.
\end{align}
The amplitude finally reads as
\begin{equation}
	\mathcal{M}
	=
	\frac{1}{4\omega_1^2\omega_2^2}
	\left(
	\mathcal{S}_0
	+
	\mathcal{S}_{\mathrm{SO}}^{(1)}
	+
	\mathcal{S}_{\mathrm{SO}}^{(2)}
	+
	\mathcal{S}_{\mathrm{SS}}
	\right)~.
\end{equation}

 \subsection{Spin-spin effective interaction in coordinate space}
 
 The localized evolution equation for the density matrix, after integrating out
 the photon degrees of freedom (photon propagators and energy denominators) and
 applying the Markov approximation (\(\tau \to t\) closure), takes the form
 \begin{align}
 	\dot{\rho}_{IJ}(\mathbf{k},t)
 	&=
 	-i q^4 V \frac{\delta^3(\Delta \mathbf x_A)}{m_A}
 	\sum_{\substack{r_A,r'_A\\ r_{B_1},r'_{B_1}\\ r_{B_2},r'_{B_2}}}
 	\int d\mathbf K_1\, d\mathbf K_2\;
 	\mathcal M(\mathbf k,\mathbf K_1,\mathbf K_2)
 	\exp\!\Big[
 	-i\mathbf K_1\!\cdot\!(\bar{\mathbf x}_{1A}-\bar{\mathbf x}_{1B})
 	-i\mathbf K_2\!\cdot\!(\bar{\mathbf x}_{2A}-\bar{\mathbf x}_{2B})
 	\Big]
 	\nonumber\\
 	&\quad\times
 	\exp\!\left[
 	-\frac{\sigma_A^2}{4}(|\mathbf K_1|^2+|\mathbf K_2|^2)
 	-\frac{\sigma_{B_1}^2}{2}|\mathbf K_1|^2
 	-\frac{\sigma_{B_2}^2}{2}|\mathbf K_2|^2
 	\right]
 	\Big[
 	\delta_{r_A i}\delta_{r_{B_1}k_1}\delta_{r_{B_2}k_2}
 	\rho^A_{r'_A j}(\mathbf k)
 	\rho^{B_1}_{r'_{B_1}l_1}(\mathbf k)
 	\rho^{B_2}_{r'_{B_2}l_2}(\mathbf k)
 	\nonumber\\
 	&\qquad\qquad\qquad\qquad\qquad
 	-
 	\delta_{j r'_A}\delta_{l_1 r'_{B_1}}\delta_{l_2 r'_{B_2}}
 	\rho^A_{i r_A}(\mathbf k)
 	\rho^{B_1}_{k_1 r_{B_1}}(\mathbf k)
 	\rho^{B_2}_{k_2 r_{B_2}}(\mathbf k)
 	\Big]~.
 \end{align}
 We now integrate over \(\mathbf K_1\) and \(\mathbf K_2\), retaining only the
 spin-spin contribution \(\mathcal S_{\mathrm{SS}}\) from
 \(\mathcal M\).
 We define the relative separations
 \begin{equation}
 	\mathbf r_1
 	=
 	\bar{\mathbf x}_{1A}-\bar{\mathbf x}_{1B}~,
 	\qquad
 	\mathbf r_2
 	=
 	\bar{\mathbf x}_{2A}-\bar{\mathbf x}_{2B}~.
 \end{equation}
  The Gaussian smearing factors are collected into effective spatial widths
 \begin{equation}
 	a_1
 	=
 	\frac{\sigma_A^2}{4}
 	+
 	\frac{\sigma_{B_1}^2}{2}~,
 	\qquad
 	a_2
 	=
 	\frac{\sigma_A^2}{4}
 	+
 	\frac{\sigma_{B_2}^2}{2}~.
 \end{equation}
  The momentum integrals involve the Fourier transform of the
 smeared Coulomb kernel
 \begin{equation}
 	F_a(r)
 	=
 	\int d\mathbf K\;
 	\frac{e^{-i\mathbf K\cdot\mathbf r}}{|\mathbf K|^2}
 	e^{-a|\mathbf K|^2}
 	=
 	\frac{1}{4\pi r}
 	\operatorname{erf}\!\left(\frac{r}{2\sqrt a}\right)~.
 \end{equation}
  Spatial derivatives of \(F_a(r)\) generate the required momentum factors
 \begin{align}
 	\nabla_i F_a(r)
 	&=
 	i
 	\int d\mathbf K\;
 	\frac{K_i}{|\mathbf K|^2}
 	e^{-i\mathbf K\cdot\mathbf r-a|\mathbf K|^2}~,
 	\\
 	\nabla_i\nabla_j F_a(r)
 	&=
 	-
 	\int d\mathbf K\;
 	\frac{K_iK_j}{|\mathbf K|^2}
 	e^{-i\mathbf K\cdot\mathbf r-a|\mathbf K|^2}~.
 \end{align} 
 Performing the momentum integrations yields the coordinate space kernel
 \begin{align}
 	T_{\mathrm{SS}}
 	&=
 	-\frac{1}{4m_{B_1}m_{B_2}}
 	(\chi_{r'_A}^\dagger\chi_{r_A})
 	\Big[
 	(\chi_{r'_{B_1}}^\dagger
 	\boldsymbol{\sigma}
 	\chi_{r_{B_1}})
 	\cdot
 	(\chi_{r'_{B_2}}^\dagger
 	\boldsymbol{\sigma}
 	\chi_{r_{B_2}})
 	\,
 	\nabla_{\mathbf r_1}F_{a_1}(r_1)
 	\!\cdot\!
 	\nabla_{\mathbf r_2}F_{a_2}(r_2)
 	\nonumber\\
 	&\qquad
 	-
 	(\chi_{r'_{B_1}}^\dagger
 	\boldsymbol{\sigma}\!\cdot\!\nabla_{\mathbf r_2}F_{a_2}(r_2)
 	\,\chi_{r_{B_1}})
 	(\chi_{r'_{B_2}}^\dagger
 	\boldsymbol{\sigma}\!\cdot\!\nabla_{\mathbf r_1}F_{a_1}(r_1)
 	\,\chi_{r_{B_2}})
 	\Big]~.
 \end{align}
The effective interaction is therefore governed by spatial derivatives of
 the smeared Coulomb kernel
 \begin{equation}
 	F_a(r)
 	=
 	\frac{1}{4\pi r}
 	\operatorname{erf}\!\left(\frac{r}{2\sqrt a}\right)~,
 \end{equation}
 which generates the tensor structure of the spin-spin interaction.
  Substituting \(T_{\mathrm{SS}}\) back into the evolution equation and using
$
 	\delta^3(\Delta\mathbf x_A)V=1
$
 gives
 \begin{align}
 	\dot{\rho}_{IJ}(\mathbf k,t)
 	&=
 	-i\frac{q^4}{m_A}
 	\sum_{\substack{r_A,r'_A\\ r_{B_1},r'_{B_1}\\ r_{B_2},r'_{B_2}}}
 	T_{\mathrm{SS}}
 	\Big[
 	\delta_{r_A i}\delta_{r_{B_1}k_1}\delta_{r_{B_2}k_2}
 	\rho^A_{r'_A j}
 	\rho^{B_1}_{r'_{B_1}l_1}
 	\rho^{B_2}_{r'_{B_2}l_2}
 	-
 	\delta_{j r'_A}\delta_{l_1 r'_{B_1}}\delta_{l_2 r'_{B_2}}
 	\rho^A_{i r_A}
 	\rho^{B_1}_{k_1 r_{B_1}}
 	\rho^{B_2}_{k_2 r_{B_2}}
 	\Big]~.
 \end{align}
 To evaluate the spin sums we use the Pauli spinor identities
$
 	\chi_r^\dagger\chi_s
 	=
 	\delta_{rs}
 $ and $
 	\chi_r^\dagger\sigma^\alpha\chi_s
 	=
 	(\sigma^\alpha)_{rs}.
$
 Hence, the mediator spin factor reduces to
$
 	\chi_{r'_A}^\dagger\chi_{r_A}
 	=
 	\delta_{r'_Ar_A}
$.
  For the term proportional to
 \(
 \delta_{r_A i}\delta_{r_{B_1}k_1}\delta_{r_{B_2}k_2}
 \),
 the mediator spin contraction gives
 \begin{equation}
 	\sum_{r'_A}
 	\delta_{r'_Ai}\rho^A_{r'_Aj}
 	=
 	\rho^A_{ij}~.
 \end{equation}
 The bath spin contractions become
 \begin{align}
 	\sum_{r'_{B_1}}
 	(\sigma^\alpha)_{r'_{B_1}k_1}
 	\rho^{B_1}_{r'_{B_1}l_1}
 	&=
 	(\sigma^\alpha\rho^{B_1})_{k_1l_1}~,
 	\\
 	\sum_{r'_{B_2}}
 	(\sigma^\beta)_{r'_{B_2}k_2}
 	\rho^{B_2}_{r'_{B_2}l_2}
 	&=
 	(\sigma^\beta\rho^{B_2})_{k_2l_2}~.
 \end{align}
  For the term proportional to
 \(
 \delta_{jr'_A}\delta_{l_1r'_{B_1}}\delta_{l_2r'_{B_2}}
 \),
 the mediator contraction again gives \(\rho^A_{ij}\), while the bath spin sums become
 \begin{align}
 	\sum_{r_{B_1}}
 	\rho^{B_1}_{k_1r_{B_1}}
 	(\sigma^\alpha)_{l_1r_{B_1}}
 	&=
 	(\rho^{B_1}\sigma^\alpha)_{k_1l_1}~,
 	\\
 	\sum_{r_{B_2}}
 	\rho^{B_2}_{k_2r_{B_2}}
 	(\sigma^\beta)_{l_2r_{B_2}}
 	&=
 	(\rho^{B_2}\sigma^\beta)_{k_2l_2}~.
 \end{align}
Finally, the master equation takes the form
 \begin{equation}
 	\dot{\rho}_{IJ}
 	=
 	i\frac{q^4}{4m_Am_{B_1}m_{B_2}}
 	T_{\alpha\beta}\,
 	\rho^A_{ij}
 	\left[
 	(\sigma^\alpha\rho^{B_1})_{k_1l_1}
 	(\sigma^\beta\rho^{B_2})_{k_2l_2}
 	-
 	(\rho^{B_1}\sigma^\alpha)_{k_1l_1}
 	(\rho^{B_2}\sigma^\beta)_{k_2l_2}
 	\right]~,
 \end{equation}
with the spatial tensor
 \begin{equation}
 	T_{\alpha\beta}
 	=
 	\delta_{\alpha\beta}\,
 	\nabla_{\mathbf r_1}F_{a_1}
 	\!\cdot\!
 	\nabla_{\mathbf r_2}F_{a_2}
 	-
 	(\nabla_{\mathbf r_2}F_{a_2})_\alpha
 	(\nabla_{\mathbf r_1}F_{a_1})_\beta~. 
 \end{equation}

\end{document}